\documentclass[12pt]{article}
\usepackage{amsfonts}
\usepackage{amsmath}
\usepackage{amssymb}
\usepackage{array}
\usepackage{bigints}
\usepackage{booktabs}
\usepackage[nosort]{cite}
\usepackage{color}
\usepackage{dsfont}
\usepackage{float}
\usepackage{framed}
\usepackage{graphicx}
\usepackage{indentfirst}
\usepackage{mathrsfs}
\usepackage{multirow}
\usepackage{setspace}
\usepackage{subfig}
\usepackage{titlesec}
\usepackage[dotinlabels]{titletoc}
\usepackage{wrapfig}
\usepackage[all]{xy}
\usepackage{young}
\usepackage[vcentermath]{youngtab}

\usepackage{hyperref}
\hypersetup{colorlinks=true}
\hypersetup{linkcolor=black}
\hypersetup{citecolor=black}
\hypersetup{urlcolor=black}

\numberwithin{equation}{section}

\usepackage[left=2.5cm,right=2.5cm,top=2.5cm,bottom=3cm]{geometry}
\titleformat{\section}{\normalfont\bfseries}{\thesection.}{4pt}{}
\titlespacing{\section}{0pt}{20pt}{6pt}

\titleformat{\subsection}{\normalfont\itshape}{\thesubsection.}{4pt}{}
\titlespacing{\subsection}{0pt}{15pt}{6pt}

\titleformat{\subsubsection}{\normalfont}{\thesubsubsection.}{4pt}{}
\titlespacing{\subsubsection}{0pt}{15pt}{6pt}

\def\ie{\begin{equation}\begin{aligned}}
\def\fe{\end{aligned}\end{equation}}

\newcommand\VRule[1][\arrayrulewidth]{\vrule width #1}

\def\hat{\widehat}

\def\half{{1 \over 2}}
\def\d{\partial}

\def\1{{\mathds 1}}

\DeclareMathOperator{\Tr}{\mathrm{Tr}}

\def\alphadot{{\dot\alpha}}

\newcommand{\Z}{{\mathbb Z}}

\newcommand{\R}{{\mathbb R}}

\def\SL{{\mathscr L}}

\def\CA{{\mathcal A}}

\def\CD{{\mathcal D}}

\def\CF{{\mathcal F}}

\def\CI{{\mathcal I}}

\def\CM{{\mathcal M}}
\def\CN{{\mathcal N}}
\def\CO{{\mathcal O}}

\def\CT{{\mathcal T}}

\def\CW{{\mathcal W}}

\DeclareFontShape{OT1}{cmr}{mx}{n}%
    {<->cmr10}{}
\newcommand{\mytitlefont}{\fontseries{mx}\selectfont}
\DeclareMathAlphabet{\titlemath}{OT1}{cmr}{mx}{n}

\begin{document}


\begin{titlepage}

\begin{center}

~\\[2cm]

{\fontsize{25pt}{0pt} \mytitlefont Higher Derivative Terms, Toroidal \\[1pt]Compactification, and Weyl Anomalies \\[2pt] in Six-Dimensional~$(2,0)$ Theories}

~\\[0.5cm]

Clay C\'{o}rdova,$^1$ Thomas T.~Dumitrescu,$^2$ and Xi Yin$^2$

~\\[0.1cm]

$^1$ {\it Society of Fellows, Harvard University, Cambridge, MA 20138, USA}

$^2$ {\it Department of Physics, Harvard University, Cambridge, MA 20138, USA}

~\\[0.8cm]

\end{center}

\noindent We systematically analyze the effective action on the moduli space of~$(2,0)$ superconformal field theories in six dimensions, as well as their toroidal compactification to maximally supersymmetric Yang-Mills theories in five and four dimensions. We present a streamlined approach to non-renormalization theorems that constrain  this effective action. The first several orders in its derivative expansion are determined by a one-loop calculation in five-dimensional Yang-Mills theory. This fixes the leading higher-derivative operators that describe the renormalization group flow into theories residing at singular points on the moduli space of the compactified~$(2,0)$ theories. This understanding allows us to compute the~$a$-type Weyl anomaly for all~$(2,0)$ superconformal theories. We show that it decreases along every renormalization group flow that preserves~$(2,0)$ supersymmetry, thereby establishing the~$a$-theorem for this class of theories. Along the way, we encounter various field-theoretic arguments for the ADE classification of~$(2,0)$ theories.

\vfill

\begin{flushleft}
May 2015
\end{flushleft}

\end{titlepage}


\tableofcontents

\section{Introduction}

In this paper, we study aspects of~$(2,0)$ superconformal field theories (SCFTs) in six dimensions, as well as their circle and torus compactifications, which lead to maximally supersymmetric Yang-Mills theories in five and four dimensions. (We will refer to them as~$\CN=2$ and~$\CN=4$ in their respective dimension.) In every case, we analyze the low-energy effective action for the massless fields on the moduli space of vacua. We explain a streamlined approach to the powerful non-renormalization theorems of~\cite{Paban:1998ea,Paban:1998mp,Paban:1998qy,Sethi:1999qv,Maxfield:2012aw}, which follow from maximal supersymmetry.\footnote{~These non-renormalization theorems were originally found by analyzing higher-derivative corrections to BFSS matrix quantum mechanics~\cite{Banks:1996vh}. By contrast with the non-renormalization theorems discussed in~\cite{Dine:1997nq}, which only require eight supercharges but use superconformal symmetry to constrain certain~$D$-terms, the ones discussed in this paper require maximal supersymmetry but also apply in theories that are not conformal.} This allows us to show that the functional form of the effective action at the first several orders in the derivative expansion is completely fixed in terms of a few coefficients. These can then be tracked along the moduli space, and across dimensions. Up to six-derivative order, all such coefficients can be determined from a one-loop calculation in five-dimensional~$\CN=2$ Yang-Mills theory, using only the standard two-derivative Lagrangian. Although this Yang-Mills Lagrangian is expected to be corrected by irrelevant operators, we show that the only such operators that could contaminate our results are inconsistent with the conformal symmetry of the six-dimensional parent theory. We also explain why it is in general not possible to reproduce the one-loop result in five dimensions from an analogous calculation in a genuinely four-dimensional~$\CN=4$ Yang-Mills theory. 

This understanding leads to a computation of the~$a$-type Weyl anomaly for all~$(2,0)$ SCFTs, in the spirit of~\cite{Maxfield:2012aw}, and a new calculation of their~$R$-symmetry anomaly, along the lines envisioned in~\cite{Intriligator:2000eq}. These papers argued that both anomalies are captured by certain six-derivative terms in the moduli-space effective action, and proposed to fix the coefficients by comparing to~$\CN=2$ or~$\CN=4$ Yang-Mills theory in five or four dimensions; they also raised several puzzles that will be addressed below. Using our results, we show that the~$a$-anomaly is strictly decreasing under all renormalization group (RG) flows that preserve~$(2,0)$ supersymmetry, thus verifying the conjectured~$a$-theorem~\cite{Cardy:1988cwa} in six dimensions for this class of flows. We also discuss several field-theoretic arguments for the ADE classification of the~$(2,0)$ theories. One of these arguments only relies on consistency conditions for the moduli-space effective action in six dimensions. 

We begin with a brief review of~$(2,0)$ SCFTs, and what is known about their anomalies, before stating our assumptions and summarizing our results in more detail. 

\subsection{Review of~$(2,0)$ Theories}

In Lorentzian signature, the~$(2,0)$ superconformal algebra in six dimensions is~$\frak{osp} (8^*|4)$, whose Bosonic subalgebra is the sum of the conformal algebra~$\frak{so}(6,2)$ and the~$R$-symmetry algebra~$\frak {sp}(4)_R = \frak{ so}(5)_R$. All well-defined local operators reside in unitary representations of~$\frak{osp} (8^*|4)$. The only such representation that describes standard free fields is the Abelian tensor multiplet, which contains the following operators:
\begin{itemize}
\item Real scalars~$\Phi^I \; (I = 1,\ldots, 5)$ in the~$\bf 5$ of~$\frak{so}(5)_R$. They satisfy~$\square \Phi^I = 0$ and have scaling dimension~$\Delta = 2$.

\item Weyl Fermions in a~$\bf 4$ of the~$\frak{so}(5,1)$ Lorentz algebra and the~$\bf 4$ of~$\frak{so}(5)_R$, subject to a symplectic Weyl reality condition. They satisfy the free Dirac equation and have scaling dimension~$\Delta = {5\over 2}$. 

\item A real, self-dual three-form~$H = *H$, which is the field strength of a two-form gauge field~$B$. Therefore~$H = dB$ is closed and co-closed, $dH = d* H= 0$, and its scaling dimension is~$\Delta = 3$. 

\end{itemize} 

Since free field theories only admit conventional relevant or marginal interactions in~$d \leq 4$ dimensions, interacting field theories in six dimensions were long thought to be impossible. Surprisingly, decoupling limits of string theory strongly suggest the existence of interacting~$(2,0)$ SCFTs in six dimensions~\cite{Witten:1995zh,Strominger:1995ac,Witten:1995em}. For a review, see~\cite{Seiberg:1997ax, Witten:2009at,Gaiotto:2010be, moorefklect} and references therein. These theories are believed to obey standard axioms of quantum field theory~\cite{Seiberg:1996vs,Seiberg:1996qx}, such as locality and the existence of a stress tensor. However, it is not yet understood how to properly formulate them, and many properties of the known interacting~$(2,0)$ SCFTs, including their existence, have been inferred from their embedding into particular string constructions. Other aspects can be analyzed more generally. For instance, it can be shown that no~$(2,0)$ SCFT possesses relevant or marginal operators that can be used to deform the theory while preserving supersymmetry, because the superconformal algebra~$\frak{osp} (8^*|4)$ does not admit unitary representations that contain such operators~\cite{ckt}.\footnote{~The same statement also applies to~$(1,0)$ SCFTs in six dimensions~\cite{ckt}. However, all six-dimensional SCFTs have relevant deformations that break supersymmetry.}

Every~$(2,0)$ SCFT~$\CT_{\frak g}$ that can be realized in string theory is locally characterized by a real Lie algebra~$\frak g = \oplus_i \frak g_i$. (Globally, there is additional data; see for instance~\cite{Witten:2009at,moorefklect,Gaiotto:2014kfa} and references therein.)  Each~$\frak g_i$ is either~$\frak u(1)$ or a compact, simple Lie algebra of ADE type. A given~$\frak g_i$ gives rise to a theory that is locally, but not necessarily globally, decoupled from the other summands. Moreover, the~$\frak u(1)$ summands are locally described by free Abelian tensor multiplets. As long as we only probe local aspects of the theory, it is therefore sufficient to focus on one~$\frak g_i$ at a time. 

Let~$\frak g$ be~$\frak u(1)$ or a compact, simple ADE Lie algebra. In flat Minkowski space~$\R^{5,1}$, the theory~$\CT_{\frak g}$ has a moduli space of vacua,
\begin{equation}
\CM_{\frak g} = \left(\R^5\right)^{r_{\frak g}} / \CW_{\frak g}~.\label{vacua}
\end{equation}
Here~$r_{\frak g}$ and~$\CW_{\frak g}$ are the rank and Weyl group of~$\frak g$. At a generic point, the low-energy dynamics on the moduli space is described by~$r_{\frak g}$ Abelian tensor multiplets valued in the Cartan subalgebra of~$\frak g$. For this reason, we will also refer to the moduli space as the tensor branch. The vacuum expectation values (vevs) of the five scalars in each tensor multiplet parametrize the~$r_{\frak g}$ different~$\R^5$ factors, which are permuted by~$\CW_{\frak g}$. These vevs spontaneously break both the conformal and the~$\frak{so}(5)_R$ symmetry. The corresponding Nambu-Goldstone (NG) Bosons are supplied by the tensor multiplet scalars. The tensor multiplets, and hence the NG Bosons, weakly interact through higher-derivative, irrelevant operators that are suppressed by powers of the vevs. 

At the boundaries of the moduli space, the low-energy dynamics is described by an interacting superconformal theory~$\CT_{\frak h}$, where~$\frak h \subset \frak g$ is a semisimple subalgebra of lower rank~$r_{\frak h} < r_{\frak g}$, as well as~$r_{\frak g} - r_{\frak h}$ Abelian tensor multiplets. The allowed subalgebras~$\frak h \subset \frak g$ are determined  by adjoint Higgsing, so that~$\frak h$ is itself a sum of compact, simple ADE Lie algebras. Therefore, the moduli space~$\CM_{\frak g}$ has the structure one would intuitively expect from a gauge theory of non-Abelian tensor multiplets in the adjoint representation of~$\frak g$. 

This intuition can be sharpened by compactifying~$\CT_{\frak g}$ on a spatial circle of radius~$R$, with supersymmetric boundary conditions. It follows from arguments in string theory that the five-dimensional effective theory, valid below the Kaluza-Klein (KK) scale~$1 \over R$, is given by maximally supersymmetric Yang-Mills theory with gauge algebra~$\frak g$ and gauge coupling~$g^2$ proportional to~$R$. Therefore its Coulomb branch exactly coincides with~\eqref{vacua}. The fact that the five-dimensional description is a weakly-coupled non-Abelian gauge theory has been essential for exploring the dynamics of~$\CT_{\frak g}$ using field-theoretic techniques. Since this description is a standard effective theory, we expect an infinite series of irrelevant operators, suppressed by powers of the cutoff set by the KK scale.\footnote{~A different point of view was advocated in~\cite{Douglas:2010iu,Lambert:2010iw,Papageorgakis:2014dma}, where it was argued that the five-dimensional Yang-Mills description could be extended beyond its regime of validity as an effective theory with a cutoff. Here we will work within standard effective field theory, and hence our results are neither in conflict with, nor shed light on, this proposal.} What is known about these corrections will be reviewed below.

\subsection{Anomalies}

Anomalies are robust observables: even in a strongly-coupled theory, they can often be computed by utilizing different effective descriptions, some of which may be weakly coupled. In conformal field theories (CFTs), we can distinguish between 't Hooft anomalies for continuous flavor symmetries (these include gravitational and mixed anomalies) and Weyl anomalies for the conformal symmetry. 

The 't Hooft anomalies for the~$\frak{so}(5)_R$ symmetry, as well as gravitational and mixed anomalies, have been computed for all known~$(2,0)$ SCFTs~$\CT_{\frak g}$. They are summarized by an anomaly eight-form~$\CI_{\frak g}$, which encodes the anomalous variation of the action in the presence of background gauge and gravity fields via the descent procedure~\cite{AlvarezGaume:1983ig,Bardeen:1984pm}. If~$\frak g$ is~$\frak u(1)$ or a compact, simple ADE Lie algebra, then (in the normalization of~\cite{Intriligator:2000eq}),
\smallskip
\begin{equation}
\mathcal{I}_{\frak{g}}=\frac{k_{\frak g}}{24}p_{2}(F_{\frak{so}(5)_R})+r_{\frak{g}}\mathcal{I}_{\frak{u}(1)}~, \qquad k_{\frak g} = h_{\frak{g}}^{\vee}d_{\frak{g}}~,
\label{anomform}
\end{equation}
\smallskip
where~$F_{\frak{so}(5)_R}$ is the field strength of the background~$R$-symmetry gauge field, while~$h^\vee_{\frak g}$ and~$d_{\frak g}$ are the dual Coxeter number and the dimension of~$\frak g$, respectively. The anomaly polynomial~$\CI_{\frak u(1)}$ for a free Abelian tensor multiplet encodes all gravitational and mixed anomalies of~$\CT_{\frak g}$, as well as a contribution to the~$R$-symmetry anomaly. (Its precise form, which can be found in~\cite{Intriligator:2000eq}, will not be needed here.) The anomaly polynomial for~$\frak g = \frak{su}(n)$ was first obtained in \cite{Harvey:1998bx}. The general formula~\eqref{anomform} was conjectured in~\cite{Intriligator:2000eq}. It was verified in~\cite{Yi:2001bz} for~$\frak g = \frak{so}(2n)$, and in~\cite{Ohmori:2014kda} for all ADE Lie algebras.

The conjecture of~\cite{Intriligator:2000eq} was in part motivated by the insight that the constants~$k_{\frak g}$ in~\eqref{anomform} appear in the low-energy effective action for the dynamical fields on the tensor branch of~$\CT_{\frak g}$, where the theory is weakly coupled. For simplicity, we will focus on rank one adjoint breaking patterns~$\frak g \rightarrow \frak h \oplus \frak u(1)$, which lead to moduli spaces described by a single Abelian tensor multiplet. The difference~$\Delta k = k_{\frak g} - k_{\frak h}$ is the coefficient of a Wess-Zumino (WZ) term, which arises at six-derivative order. This term is needed to match the irreducible 't Hooft anomaly for the spontaneously broken~$\frak{so}(5)_R$ symmetry via a contribution from the~$R$-symmetry NG Bosons. It was suggested in~\cite{Intriligator:2000eq} that~$\Delta k$ could be computed by reducing the theory on a circle and integrating out massive W-Bosons on the Coulomb branch of five-dimensional~$\CN=2$ Yang-Mills theory. By analogy with~$\CN=4$ Yang-Mills theory in four dimensions, where a WZ term is generated at one loop on the Coulomb branch~\cite{Tseytlin:1999tp}, one might expect that~$\Delta k$ only depends on~$n_W = d_{\frak g} - (d_{\frak h} + 1)$, the number of~W-Bosons that become massive upon breaking~$\frak g \rightarrow \frak h \oplus \frak u(1)$. However, it was emphasized in~\cite{Intriligator:2000eq} that this is inconsistent with the values of~$\Delta k$ and~$n_W$ for the breaking patterns~$\frak{su}(n+1) \rightarrow \frak{su}(n) \oplus \frak u(1)$ and~$\frak{so}(2n) \rightarrow \frak{su}(n) \oplus \frak u(1)$, even at large~$n$. One of our goals in this paper is to explain why the four-dimensional intuition is misleading.

CFTs in even spacetime dimensions also have Weyl anomalies, which manifest as a violation of conformal invariance in the presence of a background spacetime metric. This is quantified by the anomalous trace of the stress tensor in such a background, whose scheme-independent part takes the following general form~\cite{Deser:1993yx},
\begin{equation}
\langle T^{\mu}_{\mu}\rangle= a E_{d}+\sum_{i} c_i I_i~. \label{tracet}
\end{equation}
Here the spacetime dimension~$d$ is even, $E_d$ is the Euler density, and the~$I_i$ are local Weyl invariants of weight~$d$. The number of independent~$I_i$ depends on~$d$, e.g.~in four dimensions there is one, and in six dimensions there are three.\footnote{~In two dimensions there is no~$c$-type anomaly and the~$a$-anomaly coincides with the Virasoro central charge. As a result, it is typically denoted by~$c$, even though it multiplies the two-dimensional Euler density.}

The dimensionless constants~$a$ and~$c_i$ are well-defined observables of the CFT. They are determined by certain flat-space correlators of the stress tensor at separated points. In two and four dimensions, it was shown~\cite{Zamolodchikov:1986gt,Cardy:1988cwa,Komargodski:2011vj,Komargodski:2011xv} that unitarity RG flows interpolating between CFTs in the UV and in the IR respect the following inequality,
\begin{equation}
a_{\rm UV} > a_{\rm IR}~.
\end{equation}
Therefore, $a$ provides a quantitative measure of the number of degrees of freedom in a CFT. An analogous $a$-theorem has been conjectured~\cite{Cardy:1988cwa}, and was recently investigated~\cite{Elvang:2012st}, for RG flows in six dimensions, but a proof is currently lacking. The ability to test this conjecture is limited by the paucity of interacting six-dimensional CFTs for which~$a$ has been computed. 

In~$(2,0)$ SCFTs, the three independent~$c$-type anomalies that are present in six dimensions are believed to be proportional to a single constant~$c$~(see~\cite{Bastianelli:2000hi,Beem:2014kka} for some compelling evidence). We can therefore normalize~$c_{1,2,3} = c$ in~\eqref{tracet}. It is convenient to fix the remaining normalizations by demanding that the Weyl anomalies of a free Abelian tensor multiplet~$\CT_{\frak u(1)}$, which were computed in~\cite{Bastianelli:2000hi}, take the following simple values,
\begin{equation}
a_{\frak u(1)} = c_{\frak u(1)} = 1~.
\end{equation}

Less is known about the Weyl anomalies of interacting~$(2,0)$ theories~$\CT_{\frak g}$ with~$\frak g$ a compact, simple ADE Lie algebra. At the conformal point, $c_{\frak g}$ can be extracted from a stress tensor two-point function, while~$a_{\frak g}$ requires a four-point function. For~$\frak g = \frak{su}(n)$ and~$\frak{so}(2n)$, the leading large-$n$ behavior of~$a_{\frak g}$ and~$c_{\frak g}$ can be determined form their~$AdS_7 \times S^4$ and~$AdS_7 \times \R \mathbb{P}^4$ duals~\cite{Henningson:1998gx, Intriligator:2000eq}. Subleading corrections for the~$\frak{su}(n)$ case were suggested in~\cite{Tseytlin:2000sf,Beccaria:2014qea}, motivated by aspects of the holographic dual. A recent conjecture~\cite{Beem:2014kka}, which applies to all~$\frak g$ and passes several non-trivial consistency checks, identifies~$c_{\frak g}$ with the central charge of known chiral algebra in two dimensions,
\smallskip
\begin{equation}
c_{\frak g}=4 h_{\mathfrak{g}}^{\vee} d_{\mathfrak{g}}+r_{\mathfrak{g}}~.\label{cresult}
\end{equation}
\smallskip

A method for determining~$a_{\frak g}$ was proposed in~\cite{Maxfield:2012aw}. Following the work of \cite{Komargodski:2011vj,Komargodski:2011xv} in four dimensions, it was shown in~\cite{Maxfield:2012aw,Elvang:2012st} that~$a_{\frak g}$ appears in the effective action on the tensor branch of~$\CT_{\frak g}$, where conformal symmetry is spontaneously broken. We again focus on rank one breaking~$\frak g \rightarrow \frak h \oplus \frak u(1)$. Now the difference~$\Delta a = a_{\frak g} - \left(a_{\frak h} + 1\right)$ appears as the coefficient of a six-derivative WZ-like interaction term for the dilaton (the NG Boson of spontaneous conformal symmetry breaking), which is needed to match the~$a$-anomalies of the UV and IR theories. The authors of~\cite{Maxfield:2012aw} argued for a non-renormalization theorem that fixes~$\Delta a \sim b^2$, where~$b$ is the coefficient of a four-derivative term in the tensor-branch effective action. By compactifying the theory on~$T^2$ and tracking this four-derivative term as the torus shrinks to zero size, they argued that~$b$ could be extracted from a one-loop computation in four-dimensional~$\CN=4$ Yang-Mills theory with gauge algebra~$\frak g$. This leads to~$b \sim n_W$, the number of massive~W-Bosons. However, just as for~$\Delta k$ above, the large-$n$ asymtotics of~$a_{\frak g}$ for~$\frak g = \frak{su}(n)$ and~$\frak g = \frak{so}(2n)$, which are known from holography, imply that~$\Delta a$ cannot just depend on~$n_W$. This puzzle will also be resolved in the course of our investigation. 

\bigskip

\subsection{Assumptions}

In this paper, we will analyze the moduli space effective actions of~$(2,0)$ SCFTs and their compactifications on~$S^1$ and~$T^2$. The goal is to learn as much as possible about the interacting theories using field-theoretic arguments and a small number of assumptions. While these assumptions are currently motivated by explicit string constructions, we hope that they can ultimately be justified for all~$(2,0)$ SCFTs -- including putative theories that do not have a known string realization. (This approach to the~$(2,0)$ theories has, for instance, been advocated in~\cite{moorefklect}.) In this spirit, the arguments and conclusions of this paper only rely on the following assumption.

\begin{framed}
\noindent{\bf Assumption:} When a six-dimensional~$(2,0)$ SCFT is supersymmetrically compactified on a spatial~$S^1_R$ of radius~$R$, the effective low-energy description at energies far below the Kaluza-Klein scale~$1 \over R$ is given by a five-dimensional~$\CN=2$ Yang-Mills theory. 
\end{framed}
\noindent As was already stressed above, we expect this low-energy description to be corrected by irrelevant operators, which are suppressed by powers of the KK scale, and we make no a priori assumptions about their coefficients.

Since unitarity requires the gauge algebra~$\frak g$ in five dimensions to be a real Lie algebra comprised of~$\frak u(1)$ and compact simple summands, our assumption implies that every~$(2,0)$ theory is associated with such a Lie algebra. For the remainder of this paper, we will use~$\CT_{\frak g}$ to collectively denote all~$(2,0)$ theories that give rise to five-dimensional Yang-Mills theories with gauge algebra~$\frak g$. However, we will not assume that all such theories are the same. For instance, their five-dimensional descriptions might differ by irrelevant operators. From this point of view, it is no longer clear that a~$(2,0)$ SCFT whose five-dimensional description is a~$\frak u(1)$ gauge theory must be a free Abelian tensor multiplet, but we will show that this is indeed the case in section~3.4.  For ease of presentation, we will focus on one summand of~$\frak g$ at a time, since our results are not affected by this. Throughout the paper, $\frak g$ will therefore denote~$\frak u(1)$ or a compact, simple Lie algebra. 

We also do not input the assumption that~$\frak g$ is simply laced, even though we are considering a standard~$S^1$ compactification. (In particular, we do not turn on outer automorphism twists~\cite{Vafa:1997mh}; see~\cite{Witten:2009at,Tachikawa:2011ch} for a recent discussion.) Instead, we will allow arbitrary~$\frak g$ and derive the ADE restriction from consistency conditions in field theory. We will also review standard arguments that show why~$\CT_{\frak g}$ has a moduli space of vacua given by~\eqref{vacua}, and why the allowed breaking patterns are determined by adjoint Higgsing. 

So far we have only mentioned the gauge algebra~$\frak g$. In gauge theory, one should specify a gauge group~$G$, whose Lie algebra is~$\frak g$. This is needed to define global aspects of the theory, such as the spectrum of allowed line operators or partition functions on topologically non-trivial manifolds. In maximally supersymmetric Yang-Mills theories that descend from~$(2,0)$ SCFTs, the choice of~$G$ arises due to subtle properties of the six-dimensional parent theory (see for instance~\cite{Witten:2009at,moorefklect,Tachikawa:2013hya}). Much of our discussion only refers to the Lie algebra~$\frak g$, but we will also encounter some global issues that depend on a choice of gauge group~$G$.  

\subsection{Summary of Results}

In section~2 we consider the low-energy effective action on the moduli space of~$\CT_{\frak g}$ in flat Minkowski space~$\R^{5,1}$. We focus on rank one breaking patterns~$\frak g \rightarrow \frak h \oplus \frak u(1)$, so that the moduli space is described by a single Abelian tensor multiplet. We review the WZ-like six-derivative terms that are required for the matching of the~$R$-symmetry anomaly in~\eqref{anomform} and the Weyl~$a$-anomaly in~\eqref{tracet} between the UV and IR theories~\cite{Intriligator:2000eq,Maxfield:2012aw,Elvang:2012st}. We then give a simple proof of the non-renormalization theorem of~\cite{Maxfield:2012aw}, which implies that all terms in the effective action up to six-derivative order are controlled by a single coefficient~$b$ residing at four-derivative order. In particular, the coefficients of the six-derivative WZ terms are quadratically related to~$b$,\footnote{~As we will explain in section~2, this statement only holds if the fields in the Abelian tensor multiplet are canonically normalized.}
\begin{equation}\label{deltaka}
\Delta k = k_{\frak g} -k_{\frak h} \sim b^2~, \qquad \Delta a = a_{\frak g} - (a_{\frak h} +1) \sim b^2~.
\end{equation}
Here~$\sim$ implies equality up to model-independent constants that are fixed by supersymmetry. Our proof of the non-renormalization theorems leading to~\eqref{deltaka} only relies on results from superconformal representation theory~\cite{ckt}. We also present a complementary point of view based on scattering superamplitudes for the fields in the tensor multiplet. 

In section~3 we study the~$(2,0)$ theories~$\CT_{\frak g}$ on~$\R^{4,1} \times S^1_R$. By our assumption in section~1.3, the low-energy description is a five-dimensional~$\CN=2$ Yang-Mills theory deformed by higher-derivative operators. As in six dimensions, we use non-renormalization theorems to track the Coulomb-branch effective action from the origin, where the five-dimensional Yang-Mills description is valid, to large vevs, where it is simply related to the tensor branch effective action of the six-dimensional parent theory~$\CT_{\frak g}$. This leads to new results about both regimes. The Yang-Mills description allows us to calculate the coefficient~$b$ appearing in~\eqref{deltaka} by integrating out W-Bosons at one loop, and via~\eqref{deltaka} also the coefficients of the six-dimensional WZ terms. (This is possible despite the fact -- emphasized in~\cite{Intriligator:2000eq} -- that the~$R$-symmetry WZ term vanishes when it is reduced to five dimensions.) Conversely, the conformal invariance of~$\CT_{\frak g}$ forces the leading possible higher-derivative operators at the origin of the five-dimensional Coulomb branch to vanish. Finally, matching the massive~$\half$-BPS states on the Coulomb and tensor branches leads to the requirement that~$\frak g$ be simply laced.  

In section~4, we combine the computation of~$b$ from section~3 with the relations~\eqref{deltaka} obtained in section~2 to compute the Weyl anomaly~$a_{\frak g}$ and the~$R$-symmetry anomaly~$k_{\frak g}$ for all~$(2,0)$ theories~$\CT_{\frak g}$. For the~$a$-anomaly, we find 
\begin{equation}
a_{\frak g}=\frac{16}{7}h_{\mathfrak{g}}^{\vee} d_{\mathfrak{g}}+r_{\mathfrak{g}}~, \label{aresult}
\end{equation}
in agreement with previous large-$n$ results for~$\frak g = \frak{su}(n)$ and~$\frak g = \frak{so}(2n)$ from holography. We also show that the~$a$-anomaly decreases under all RG flows that preserve~$(2,0)$ supersymmetry, in agreement with the conjectured~$a$-theorem~\cite{Cardy:1988cwa}. As we will see, the positivity of~$\Delta a$ for all such flows essentially follows from~\eqref{deltaka}. 

For the~$R$-symmetry anomaly~$k_{\frak g}$ we recover~\eqref{anomform}, which was derived in~\cite{Ohmori:2014kda} by considering a one-loop exact Chern-Simons term in five dimensions that involves dynamical and background fields. By contrast, we access~$\Delta k$ through a six-derivative term for the dynamical fields. The coefficient~$\Delta k$ is quantized, because it multiplies a WZ term in the tensor-branch effective action. We show that this quantization condition can always be violated when~$\frak g$ is not simply laced. This constitutes an alternative argument for the ADE restriction on~$\frak g$. 

Both our result~\eqref{aresult} for the~$a$-anomaly, and the conjectured formula~\eqref{cresult} for the~$c$-anomaly are linear combinations of~$h^\vee_{\frak g} d_{\frak g}$ and~$r_{\frak g}$, which determine the anomaly eight-form in~\eqref{anomform}. In fact, once such a relationship between the independent Weyl and 't Hooft anomalies is assumed, \eqref{cresult} and~\eqref{aresult} can be obtained by fitting to the known anomalies of a free Abelian tensor multiplet and one reliable large-$n$ example from holography. As in four dimensions~\cite{Anselmi:1997am}, linear relations between Weyl and 't Hooft anomalies are captured by anomalous stress-tensor supermultiplets, which contain both the anomalous divergence of the~$R$-current and the anomalous trace of the stress tensor. For six-dimensional SCFTs, these anomaly multiplets are currently under investigation~\cite{ctkii}.

In section 5 we consider~$(2,0)$ SCFTs~$\CT_{\frak g}$ on~$\R^{3,1} \times T^2$, where~$T^2 = S^1_R \times S^1_r$ is a rectangular torus of finite area~$A = R r$. We describe their moduli spaces of vacua, which depend on a choice of gauge group~$G$, and the singular points at which there are interacting~$\CN=4$ theories. In addition to the familiar theory with gauge group~$G$, which resides at the origin, there are typically additional singular points at finite distance~$\sim A^{-\half}$ (sometimes with a different gauge group), which move to infinite distance when the torus shrinks to zero size.  We illustrate these phenomena in several examples and explain the underlying mechanism. As before, we use non-renormalization theorems to determine the four-derivative terms in the Coulomb-branch effective action via a one-loop calculation in five-dimensional~$\CN=2$ Yang-Mills theory, which now includes a sum over KK modes. In general, the result cannot be interpreted as arising from a single~$\CN=4$ theory in four dimensions. We also determine the leading higher-derivative operators that describe the RG flow into the different interacting~$\CN=4$ theories on the moduli space. 

Appendix~A summarizes aspects of scattering superamplitudes in six and five dimensions, which provide an alternative approach to the non-renormalization theorems discussed in this paper (see especially section~2.3).

\section{The Tensor Branch in Six Dimensions}

In this section we analyze the low-energy effective Lagrangian~$\SL_{\text{tensor}}$ on the tensor branch of a~$(2,0)$ theory~$\CT_{\frak g}$ in six-dimensional Minkowski space~$\R^{5,1}$. For simplicity, we focus on branches of moduli space described by a single Abelian tensor multiplet, which arise from breaking patters of the form~$\frak g \rightarrow \frak h \oplus \frak u(1)$. Here~$\frak h$ is a sum of compact simple Lie algebras that is obtained by deleting a single node in the Dynkin diagram of~$\frak g$, i.e.~by adjoint Higgsing. We review what is known about~$\SL_{\text{tensor}}$  on general grounds, before turning to a systematic discussion of the constraints that follow from supersymmetry. 

\subsection{General Properties}

We consider branches of moduli space that are parametrized by the vevs~$\langle \Phi^I \rangle$ of the five scalars in a single Abelian tensor multiplet. At low energies, the fields in the tensor multiplet become free, i.e.~they satisfy the free equations of motion reviewed in section~1.1. Naively, these are summarized by a quadratic Lagrangian, 
\begin{equation}
\SL_{\text{free}} = -\half \sum_{I = 1}^5 (\d_\mu\Phi^I)^2 - \half H \wedge * H + (\text{Fermions})~, \label{freetensoract}
\end{equation}
where the signs are due to the fact that we are working in Lorentzian signature~$- + \cdots +$.  However, the fact that~$H$ is self-dual implies that~$H\wedge *H = 0$. While this is not a problem classically, where the self-duality constraint can be imposed on the equations of motion, defining the quantum theory of a free self-dual three-form requires some sophistication. Nevertheless, it is well understood  (see for instance~\cite{Witten:2007ct,Witten:2009at,moorefklect} and references therein). Below we will deform~$\SL_{\text{free}}$ by adding higher-derivative operators constructed out of the fields in the tensor multiplet. These cause no additional complications beyond those that there are already present at the two-derivative level. With this in mind, we will use~$\SL_{\text{free}}$ to denote the theory of a free Abelian tensor multiplet. 

The vevs~$\langle\Phi^I\rangle$ spontaneously break both the conformal symmetry and the~$R$-symmetry. Since~$\Phi^I$ is a vector of~$\frak{so}(5)_R$, the~$R$-symmetry is broken to~$\frak{so}(4)_R$. It is convenient to introduce radial and transverse variables,
\begin{equation}\label{ngBosons}
\Psi = \left(\sum_{I = 1}^5 \Phi^I \Phi^I\right)^\half~, \qquad \hat \Phi^I = {\Phi^I \over \Psi}~.
\end{equation}
The field~$\Psi$ has dimension two, while the~$\hat \Phi^I$ are dimensionless. Therefore~$\langle \Psi \rangle$ is the only dimensionful parameter, which sets the scale of conformal and~$R$-symmetry breaking. The fluctuations of~$\Psi$ describe the dilaton, the NG Boson of conformal symmetry breaking, and the fluctuations of the transverse fields~$\hat \Phi^I$ describe the four NG Bosons that must arise upon breaking~$\frak{so}(5)_R \rightarrow \frak{so}(4)_R$. Note that~$\sum_{I = 1}^5 \hat \Phi^I \hat \Phi^I = 1$, so that the~$\hat \Phi^I$ describe a unit~$S^4 = SO(5)_R/SO(4)_R$\,. 

Upon activating~$\langle \Phi^I \rangle$, some degrees of freedom in~$\CT_{\frak g}$ acquires masses of order~$\sqrt{\langle \Psi\rangle}$. The remaining massless degrees of freedom are the interacting theory~$\CT_{\frak h}$ and the Abelian tensor multiplet containing the NG Bosons~\eqref{ngBosons}. It follows from Goldstone's theorem that the theories at the origin and on the tensor branch decouple at very low energies, and moreover that the multiplet of NG Bosons becomes free. We will focus on an effective Lagrangian~$\SL_{\text{tensor}}$ for the Abelian tensor multiplet. Integrating out the massive degrees of freedom at the scale~$\sqrt{\langle \Psi\rangle}$ induces weak, higher-derivative interactions for the NG Bosons and their superpartners.\footnote{~We follow the standard rules for counting derivatives in supersymmetric theories: $\d_\mu$ and~$H$ both have weight~$1$, Fermions (including the supercharges) have weight~$\half$, and the scalars~$\Phi^I$ have weight~$0$.} Schematically, 
\begin{equation}\label{ltensor}
\SL_{\text{tensor}} = \SL_{\text{free}} + \sum_i f_i(\Phi^I) \CO_i~,
\end{equation}
where~$\CO_i$ is a higher-derivative operator of definite~$R$-charge and scaling dimension that is constructed out of fields in the Abelian tensor multiplet. The higher-derivative interactions are constrained by (non-linearly realized) conformal and~$R$-symmetry, as well as~$(2,0)$ supersymmetry. For instance, every~$\CO_i$ in~\eqref{ltensor} is multiplied by a scale-invariant coefficient function~$f_i(\Phi^I)$ of the moduli fields, so that their product is marginal and~$\frak{so}(5)_R$ invariant. If we expand the~$\Phi^I$ in fluctuations around their vevs~$\langle \Phi^I\rangle$, then~\eqref{ltensor} reduces to a standard effective Lagrangian with irrelevant local operators suppressed by powers of the cutoff~$\sqrt{\langle\Psi\rangle}$. Integrating out massive fields at the scale~$\sqrt{ \langle \Psi\rangle}$ also leads to irrelevant interactions that couple the NG Bosons and their superpartners to the interacting SCFT~$\CT_{\frak h}$ at the origin. Below, we will comment on why such couplings will not play a role in our discussion. 

The constraints of non-linearly realized conformal symmetry for the self-interactions of the dilaton~$\Psi$ were analyzed in~\cite{Elvang:2012st} (see also
\cite{Maxfield:2012aw,Elvang:2012yc}). The leading dilaton self-interactions arise at four-derivative order and are controlled by a single dimensionless coupling~$b$,\footnote{~Our coupling~$b$ should not be confused with a similar coupling that appears in~\cite{Elvang:2012st}. In particular, our~$b$ is dimensionless, while the one in~\cite{Elvang:2012st} is not. They are related by~$b_{\text{us}} = b_{\text{them}} / 4 \langle \Psi \rangle$.}
\begin{equation}\label{fourd}
b \, {(\d \Psi)^4 \over \Psi^3} \subset\SL_{\text{tensor}}~.
\end{equation}
Note that this term is not invariant under rescaling~$\Psi$, and hence this definition of~$b$ is tied to the canonically normalized kinetic terms in~\eqref{freetensoract}. In this normalization, $b$ controls the on-shell scattering amplitude of four dilatons at tree level~\cite{Elvang:2012st}. It follows from a dispersion relation for this amplitude that~$b \geq 0$, and that~$b = 0$ if and only if the dilaton is completely non-interacting~\cite{Adams:2006sv}, just as in the proof of the four-dimensional~$a$-theorem~\cite{Komargodski:2011vj,Komargodski:2011xv}.

At six-derivative order, conformal symmetry requires a very particular interaction term for the dilaton. Schematically, it is proportional to
\begin{equation}\label{awzw}
\Delta a \, {(\d \Psi)^6 \over \Psi^6} \subset \SL_{\text{tensor}}~, \qquad \Delta a = a_{\frak g} - \left(a_{\frak h} + 1\right)~. 
\end{equation}
Again, it is convenient to define it using tree-level dilaton scattering amplitudes~\cite{Elvang:2012st}. The term in~\eqref{awzw} is required by anomaly matching between the UV theory~$\CT_{\frak g}$ and the IR theory consisting of~$\CT_{\frak h}$ and an Abelian tensor multiplet. The~$a$-anomaly is, in a sense, irreducible and non-Abelian. It therefore requires a non-trivial WZ-like term for the dynamical dilaton, even if the background metric is flat~\cite{Komargodski:2011vj,Komargodski:2011xv}. 

A similar WZ term for the~NG Bosons~$\hat \Phi^I$ is required by anomaly matching for the~$\frak{so}(5)_{R}$ symmetry~\cite{Intriligator:2000eq}. As is typical of such a term, it is convenient to write it as an integral over a seven-manifold~$X_7$ that bounds spacetime. Under suitable conditions (explained in~\cite{Intriligator:2000eq}) we can extend the NG fields to a map~$\hat \Phi: X_{7}\rightarrow S^{4}$ and pull back the unit volume form~$\omega_4$ on~$S^4$ (i.e.~$\int_{S^4} \omega_4 = 1$) to define the a three-form~$\Omega_3$ via
\begin{equation}
d \Omega_3 = \hat \Phi^*(\omega_4)~.
\end{equation}
The WZ term of~\cite{Intriligator:2000eq} can then be written as follows,
\begin{equation}\label{kenwzw}
\frac{\Delta k}{6}\int_{X_{7}}\Omega_{3}\wedge d \Omega_{3} \subset \SL_{\text{tensor}}~, \qquad \Delta k = k_{\frak g} - k_{\frak h}~.
\end{equation}
This term is needed to match the irreducible~$\frak{so}(5)_R$ anomaly~$k$ in~\eqref{anomform} between the UV and IR theories. Requiring the six-dimensional action to be well defined leads to a quantization condition~\cite{Intriligator:2000eq},
\begin{equation}\label{wzwquant}
\Delta k \in 6 \Z~.
\end{equation}
The presence of the term~\eqref{kenwzw} and the quantization condition~\eqref{wzwquant} are general requirements, which do not rely on the known answer~\eqref{anomform} for~$k_{\frak g}$ in~ADE-type~$(2,0)$ theories. (This will play an important role in section~4.3.) Since the three-form~$\Omega_3$ only depends on the scalars~$\hat \Phi^I$, it must contain three derivatives. Therefore the integrand in~\eqref{kenwzw} is a seven-derivative term integrated over~$X_7$, which leads to a conventional six-derivative term in spacetime (albeit one that is not manifestly~$\frak{so}(5)_R$ invariant). This term gives rise to interactions involving at least seven NG Bosons.\footnote{~The same phenomenon occurs in the chiral Lagrangian for low-energy QCD: a WZ term arises at four-derivative order, and it describes an interaction involving two K mesons and three pions~\cite{Witten:1983tw}.}

In this paper, we will focus on the four- and six-derivative terms in~$\SL_{\text{tensor}}$ and their relation to anomalies via the WZ terms in~\eqref{awzw} and~\eqref{kenwzw}. As was reviewed above, these terms control the tree-level scattering amplitudes for the Abelian tensor multiplet up to and including~$\CO(p^6)$ in the momentum expansion. At higher orders in the derivative expansion, the discussion becomes more involved. On the one hand there are additional terms in~$\SL_{\text{tensor}}$, which are increasingly less constrained. On the other hand, it is no longer legitimate to ignore the interaction between the Abelian tensor multiplet and the interacting SCFT~$\CT_{\frak h}$ at the origin. Such interactions generally contribute non-analytic terms to the scattering amplitudes of tensor-branch fields that reflect the interacting massless degrees of freedom in~$\CT_{\frak h}$. However, just as in four dimensions~\cite{Komargodski:2011xv,Luty:2012ww}, these effects do not contaminate the contribution of the WZ terms at~$\CO(p^6)$, or the terms at lower order.\footnote{~We thank~Z.~Komargodski for a useful discussion about this issue.}

It is natural to ask what massive degrees of freedom should be integrated out at the scale~$\sqrt{\langle \Psi\rangle}$ in order to generate the interaction terms in~\eqref{fourd}, \eqref{awzw}, and~\eqref{kenwzw}, as well as other higher-derivative terms. In maximally supersymmetric Yang-Mills theories these are the W-Bosons and their superpartners, which become massive on the Coulomb branch. String constructions suggest that the analogous objects in~$(2,0)$ theories are dynamical strings~(see for instance~\cite{Gaiotto:2010be,moorefklect} for a review) -- we will refer to them as W-strings. On the tensor branch, the supersymmetry algebra is modified to include a brane charge~$Z_\mu^I$ proportional to~$\langle \Phi^I\rangle$.\footnote{~The supersymmetry algebra also admits another brane charge, which is associated with~$\half$-BPS three-branes, i.e.~objects of codimension two. We will not discuss them here. } This brane charge is activated by~$\half$-BPS strings, whose tension is therefore proportional to~$\langle \Psi\rangle$. On the tensor branch, such strings arise as dynamical solitons, which can be identified with the W-strings~\cite{Intriligator:2000eq}, and act as sources for  the two-form gauge field~$B$ in the tensor multiplet.  This in turn explains several features of~$\SL_{\text{tensor}}$. For instance, \eqref{wzwquant} can be interpreted as a Dirac quantization condition for these strings~\cite{Intriligator:2000eq}. It was also suggested in~\cite{Intriligator:2000eq} that the interaction terms in~$\SL_{\text{tensor}}$ might arise by integrating out the W-strings. This intuition can be made precise, and even quantitative, by compactifying the theory to five-dimensions, as we will see in section~3.

\subsection{Non-Renormalization Theorems}

We will now discuss the constraints on~$\SL_{\text{tensor}}$ that follow from supersymmetry. In particular, we will derive a strong form of the non-renormalization theorems in~\cite{Maxfield:2012aw}. As in related work on maximally supersymmetric Yang-Mills theories~\cite{Paban:1998ea,Paban:1998mp,Paban:1998qy,Sethi:1999qv}, these non-renormalization theorems were originally obtained by examining special properties of particular multi-Fermion terms in the moduli-space effective action, which imply differential equations for some of the coefficient functions~$f_i(\Phi^I)$ in~\eqref{ltensor}. Here we present a simple, general approach to these non-renormalization theorems, which elucidates their essentially group-theoretic origin. The discussion proceeds in two steps:

\begin{itemize}

\item[1.)] We expand all coefficient functions~$f_i(\Phi^I)$ in~\eqref{ltensor} in fluctuations around a fixed vev,
\begin{equation}\label{fexp}
\Phi^I = \langle \Phi^I \rangle + \delta \Phi^I~, \qquad f_i(\Phi^I) =  f_i |_{\langle \Phi\rangle} + \d_I f_i |_{\langle \Phi\rangle} \, \delta \Phi^I+ \half \d_I \d_J f_i |_{\langle \Phi\rangle} \,  \delta \Phi^I\delta \Phi^J+ \,  \cdots~.
\end{equation}
This reduces every interaction term~$f_i(\Phi^I) \CO_i$ in~\eqref{ltensor} to an infinite series of standard local operators, multiplied by suitable powers of~$\langle \Phi^I\rangle$. These local operators are constructed out of fields in the free Abelian tensor multiplet described by~$\SL_{\text{free}}$, and hence they must organize themselves into conventional (irrelevant) supersymmetric deformations of~$\SL_{\text{free}}$. 

\item[2.)] If certain operators that arise by expanding a certain coefficient function~$f_i(\Phi^I)$ cannot be identified with any supersymmetric deformation of~$\SL_{\text{free}}$, then~$f_i(\Phi^I)$ satisfies a non-renormalization theorem. This step requires a complete understanding of all supersymmetric local operators that can be constructed in the theory described by~$\SL_{\text{free}}$. 

\end{itemize}

\noindent We will now demonstrate this logic by constraining the higher-derivative terms in~$\SL_{\text{tensor}}$. However, note that the method is very general. In particular, the theory around which we expand need not be free. The main simplification is that expanding the coefficient functions as in~\eqref{fexp} leads to the problem of classifying conventional supersymmetric deformations, which can be addressed by a variety of methods.  Some applications of our approach to moduli-space non-renormalization theorems have already appeared in~\cite{Wang:2015jna,Lin:2015ixa}.

In the present example, we can use the fact that~$\SL_{\text{free}}$ is a (free) SCFT to invoke a classification of supersymmetric deformations that can be obtained using superconformal representation theory~\cite{ckt}. However, any other method of classifying supersymmetric deformations is equally valid. Below, we will also mention approaches based on scattering superamplitudes, as well as superspace. It follows from results in~\cite{ckt} that all Lorentz-scalar supersymmetric deformations of a single, free Abelian tensor multiplet fall into two classes,\footnote{~Using two or more Abelian tensor multiplets, it is also possible to construct a six-derivative~$1 \over 4$-BPS deformation~\cite{ckt}.} which we will refer to as~$F$-terms and~$D$-terms:
\begin{itemize}
\item $F$-term deformations are schematically given by
\begin{equation}\label{fterm}
\SL_F = Q^8 \left(\Phi^{(I_1} \cdots \Phi^{I_n)} - (\text{traces})\right)~, \qquad (n \geq 4)~.
\end{equation}
The operator~$\Phi^{(I_1} \cdots \Phi^{I_n)} - (\text{traces})$ is~$\half$-BPS, which is why~$\SL_F$ only involves eight supercharges. It is therefore a four-derivative term. The~$R$-symmetry indices of the supercharges and the scalars are contracted so that~$\SL_F$ transforms as a traceless symmetric~$(n-4)$-tensor of~$\frak{so}(5)_R$. The restriction~$n\geq 4$ is due to the fact that~$\SL_F$ vanishes when~$n \leq 3$. 

\item $D$-term deformations take the form
\begin{equation}\label{dterm}
\SL_D = Q^{16} \CO~,
\end{equation}
where~$\CO$ is a Lorentz scalar. Such terms always contain at least eight derivatives. 
\end{itemize}

\smallskip

As a simple illustration of our method, we use it to demonstrate the well-known non-renormalization of the kinetic terms. (As explained in section~2.1, this result is also required by conformal symmetry.) In the presence of non-trivial coefficient functions, expanding the kinetic terms leads to two-derivative interactions involving three or more fields in the Abelian tensor multiplet. For instance, 
\begin{equation}\label{kinnrt}
f_2(\Phi^I) (\d\Phi)^2 \rightarrow \left(f_2 |_{\langle \Phi\rangle} + \d_I f_2 |_{\langle \Phi\rangle} \, \delta \Phi^I + \cdots \right)  (\d\Phi)^2~,
\end{equation}
where the ellipsis denotes terms containing additional powers of~$\delta \Phi^I$. The constant~$\d_I f_2 |_{\langle \Phi\rangle}$ multiplies a two-derivative interaction of three scalars. However, the  allowed deformations~\eqref{fterm} and~\eqref{dterm} require at least four derivatives, and hence~$\d_I f_2 |_{\langle \Phi\rangle} = 0$. Since this holds at every point~$\langle \Phi^I\rangle$, we conclude that~$f_2(\Phi^I)$ cannot depend on the moduli and must in fact be a constant. Analogously, the other fields in the tensor multiplet must also have moduli-independent kinetic terms.

This reasoning straightforwardly extends to the higher derivative terms in~$\SL_{\text{tensor}}$, which first arise at four-derivative order. (There are no supersymmetric deformations of~$\SL_{\text{free}}$ with only three derivatives.) Consider, for instance,  the following term,
\begin{equation}\label{f4expand}
f_4(\Phi^I) (\d \Psi)^4 \rightarrow \left(f_4 |_{\langle \Phi\rangle} + \d_I f_4 |_{\langle \Phi\rangle} \, \delta \Phi^I+ \half \d_I \d_J f_4 |_{\langle \Phi\rangle} \, \delta \Phi^I\delta \Phi^J+ \cdots\right) (\d \Psi)^4~.
\end{equation}
The constants~$f_4 |_{\langle \Phi\rangle}$ and~$\d_I f_4 |_{\langle \Phi\rangle}$ multiply four-derivative interactions involving four and five fields. The former is~$R$-symmetry invariant, while the latter transforms as a vector of~$\frak{so}(5)_R$. These terms therefore arise as~$F$-terms~\eqref{fterm} built on~$n = 4$ and~$n=5$ scalars, respectively. The same is true for the traceless part of~$ \d_I \d_J f_4 |_{\langle \Phi\rangle}$, which multiplies a four-derivative interaction containing six fields and can arise from an~$F$-term with~$n = 6$. However, the trace~$\delta^{IJ} \d_I \d_J f_4 |_{\langle \Phi\rangle}$ multiplies an interaction with four derivatives and six fields that is invariant under the~$R$-symmetry. Neither an~$F$-term nor a~$D$-term can give rise to such an interaction. Therefore, it must vanish at every  point~$\langle \Phi^I \rangle$, and this implies that~$f_4 (\Phi^I)$ is a harmonic function, 
\begin{equation}\label{harmonic}
\delta^{IJ} \d_I \d_J f_4(\Phi^K) = 0~.
\end{equation}
Since~$f_4(\Phi^I)$ multiplies~$(\d \Psi)^4$, it follows from~$\frak{so}(5)_R$ invariance that~$f_4(\Phi^I)$ can only depend on~$\Psi$. Together with~\eqref{harmonic}, this implies that
\begin{equation}
f_4(\Phi^I)=\frac{b}{\Psi^{3}}~.\label{fsolve}
\end{equation}
A dimensionful constant term in~$f_4(\Phi^I)$ is disallowed by scale invariance. This precisely reproduces the dilaton interaction in~\eqref{fourd}. Note that~$f_4(\Phi^I) \sim {1 \over \Psi^3}$ also follows from conformal symmetry, without appealing to~\eqref{harmonic}. However, the argument that led to~\eqref{harmonic} immediately generalizes to all other coefficient functions that arise at four-derivative order, even if they multiply more complicated operators. (For instance, some of them transform in non-trivial~$R$-symmetry representations.) Therefore all of these functions are harmonic. Imposing~$R$-symmetry and scale invariance fixes their functional form up to a multiplicative constant. This was explicitly shown in~\cite{Maxfield:2012aw} for terms involving eight Fermions.

In order to see why all of these constants are in fact determined a single overall coefficient, we evaluate the coefficient functions at a fixed vev~$\langle \Phi^I\rangle$ and drop the fluctuations~$\delta \Phi^I$. All of these terms involve exactly four fields and four derivatives (at leading order in the fluctuations, scalars without derivatives only contribute vevs), and hence they must all arise from the same~$R$-symmetry invariant~$F$-term~\eqref{fterm} built on~$n = 4$ scalars. Thus, there is a single supersymmetric invariant that governs all four-derivative terms in~$\SL_{\text{tensor}}$, which are therefore controlled by a single independent coefficient. We choose it to be the constant~$b$ in~\eqref{fourd} and~\eqref{fsolve}. 

The preceding discussion of the four-derivative terms in~$\SL_{\text{tensor}}$, viewed as a deformation of~$\SL_{\text{free}}$ by local operators, was at leading order in this deformation. There are several ways to understand the effects of the deformation at higher orders: 

\begin{itemize}
\item In an on-shell approach (see e.g.~\cite{Maxfield:2012aw}), where the supersymmetry transformations~$\delta_{\text{free}}$ of~$\SL_{\text{free}}$ only close on its equations of motion, the four-derivative terms~$\SL_4$ deform the transformations to~$\delta = \delta_{\text{free}} + \delta_4$, where
\begin{equation}\label{ldeformi}
\delta_4 \SL_{\text{free}} \sim \delta_{\text{free}} \SL_4~. 
\end{equation}
Therefore, the derivative-scaling of~$\delta_4$ is~$5 \over 2$. Since every term in~$\SL_4$ is proportional to~$b$, we also have~$\delta_4 \sim b$. At the next order, the action of~$\delta_4$ on~$\SL_4$ can only be cancelled by adding a new term~$\SL_6$ to the Lagrangian, so that
\begin{equation}\label{ldeformii}
\delta_4 \SL_4 \sim \delta_0 \SL_6~.
\end{equation}
We conclude that~$\SL_6$ is a six-derivative term, which is completely determined by~$\delta_4$ and~$\SL_4$. In particular, every coefficient in~$\SL_6$ must be proportional to~$b^2$. 

\item We can determine the effects of~$\SL_4$ in conformal perturbation theory around~$\SL_{\text{free}}$. At second order, the OPE~$\SL_4(x) \SL_4(y)$ between two insertions of the deformation can contain a contact term~$\SL_6(x)\delta(x-y)$ that is required by the supersymmetry Ward identities. This contact term behaves like a tree-level insertion of~$\SL_6$, which can therefore be viewed as a term in the classical action.\footnote{~This is analogous to the seagull term~$A^\mu A_\mu |\phi|^2$ in scalar electrodynamics, which is required by current conservation and can be viewed as arising from a contact term in the OPE of two currents.} Since~$\SL_6$ arises by fusing~$\SL_4$ with itself, we conclude that~$\SL_6$ is a six-derivative term proportional to~$b^2$.\footnote{~More explicitly, the only way to generate a~$\delta$-function is by applying the equations of motion to a Wick-contraction of free fields in~$\SL_4(x)$ and~$\SL_4(y)$. This reduces the derivative order by two, which implies that~$\SL_6$ must contain six derivatives. }

\item In an approach based on scattering amplitudes, the relation of~$\SL_6$ and~$\SL_4$ can be understood through factorization. This will be discussed in section~2.3. 

\end{itemize}

\noindent All three points of view show that the four-derivative terms induce terms at six-derivative order, whose coefficients are fixed by supersymmetry and proportional to~$b^2$. In general, there could also be new supersymmetric invariants that arise at six-derivative order, with independent coefficients. However, neither the~$F$-terms~\eqref{fterm} nor the~$D$-terms~\eqref{dterm} contribute at that order, so that the six-derivative terms are completely determined by the four-derivative ones, and in particular by the coefficient~$b$. 

Since both~$\Delta a$ in~\eqref{awzw} and~$\Delta k$ in~\eqref{kenwzw} are coefficients of six-derivative terms, we conclude that they are both proportional to~$b^2$, with model-independent proportionality factors that are determined by supersymmetry. In principle these constants can be fixed by carefully working out the supersymmetry relations between the four- and six-derivative terms in~$\SL_{\text{tensor}}$. Instead, we will determine them using a reliable example. For instance, at large~$n$ the dilaton effective action for the breaking pattern~$\frak{su}(n+1) \rightarrow \frak{su}(n) \oplus \frak u(1)$ is given by the DBI action on a probe brane in~$AdS_7$. By studying this action, the authors of~\cite{Elvang:2012st} found the following relationship,\footnote{~Our normalization of the~$a$-anomaly differs from that in~\cite{Elvang:2012st}. They are related by~$a_{\text{us}} = (9216 \pi^3 /7) a_{\text{them}}$.}
\begin{equation}\label{absq}
\Delta a = \frac{98304 \, \pi^{3}}{7} \,b ^{2}~.
\end{equation}
Having fixed the constant of proportionality in this example, we conclude that~\eqref{absq} holds for all~$(2,0)$ theories and breaking patterns, because of supersymmetry.  We can similarly use the known large-$n$ behavior of~$k_{\frak{su}(n)}$ to fix
\begin{equation}\label{kbsq}
\Delta k = 6144 \pi^3 \, b^2~.
\end{equation}

\subsection{The Amplitude Point of View}

The non-renormalization theorems discussed above can also be understood by considering on-shell scattering amplitudes in the Abelian theory on the tensor branch. Since this theory is free at low energies, it is sufficient to consider tree-level amplitudes, which are meromorphic functions of the external momenta. One advantage of this approach is that supersymmetry acts linearly on one-particle asymptotic states, so that the supersymmetry Ward identities constitute a set of linear relations on the set of~$n$-point amplitudes for any given~$n$. By contrast, in an on-shell Lagrangian approach, the supersymmetry transformations of the fields are typically deformed, as discussed around~\eqref{ldeformi} and~\eqref{ldeformii}. Moreover, on-shell scattering amplitudes do not suffer from ambiguities due to field redefinitions. 

In the amplitude picture, the role of supersymmetric local operators that can be used to deform~$\SL_{\text{free}}$ is played by supervertices. These are local superamplitudes without poles that satisfy the supersymmetry Ward identities. Every supervertex corresponds to a possible first-order deformation of~$\SL_{\text{free}}$ that preserves supersymmetry. The construction of supervertices is particularly simple in the spinor helicity formalism. For the cases of interest in this paper, this formalism is reviewed in appendix A. The basic strategy is to split the 16 Poincar\'e supercharges into 8 supermomenta $Q$ and 8 superderivatives $\overline{Q}$. For instance, the four-point, four-derivative supervertex describing the supersymmetric completion of~$H^4 + (\partial\Phi)^4+\cdots$ can be written as a Grassmannian delta function~$\delta^8(Q)$. In this subsection, we will schematically denote four- and six-derivative interactions by~$H^4$ and~$H^6$, respectively.

At any point on the tensor branch, the four-derivative supervertex~$\delta^8(Q)$ is multiplied by the coefficient function~$f_4(\Phi^I)$ discussed in the previous subsection. Expanding~$f_4(\Phi^I)$ in fluctuations~$\delta\Phi^I$ of the scalar fields, as in~\eqref{fexp}, leads to soft limits of amplitudes with extra scalar emissions. Such amplitudes may or may not admit a local supersymmetric completion, as a supervertex wihtout poles. If there is no such supervertex, the corresponding soft scalar amplitude must belong to a nonlocal superamplitude, which is completely determined by the residues at its poles. This factorization relation amounts to a differential equation for the coefficient function~$f_4(\Phi^I)$, which leads to a non-renormalization theorem.

As explained in appendix A, the four-derivative, $n$-point supervertex~$\delta^8(Q)$ corresponds to a coupling of the form~$(\delta\Phi^+)^{n-4} H^4$ in~$\SL_{\text{tensor}}$. Here~$\delta\Phi^+$ is the highest-weight component in the~$\frak{so}(5)_R$ multiplet of the scalar fluctuations. All other supervertices at this derivative order are obtained from $\delta^8(Q)$ by an $\mathfrak{so}(5)_R$ rotation. In particular, the set of $n$-point supervertices at four-derivative order transform as rank $(n-4)$ symmetric traceless tensors of~$\mathfrak{so}(5)_R$. This is the amplitude version of the classification for~$F$-term deformations in~\eqref{fterm}. The absence of a supervertex that contains the~$\mathfrak{so}(5)_R$ singlet~$\delta_{IJ} \delta\Phi^I \delta\Phi^J H^4$ then leads to the requirement~\eqref{harmonic} that the coefficient function~$f_4(\Phi^I)$ be harmonic. Note that there does not exist any amplitude that contains the component vertex~$\delta\Phi^I \delta\Phi^I H^4$, since its supersymmetric completion as a six-point superamplitude would have to factorize though lower-point supervertices. However, the leading supervertices~$\delta^8(Q)$ arise at four-derivative order, so that a four-derivative amplitude cannot factorize through a pair of such vertices. 

\begin{figure}[htb]
\centering
\includegraphics[scale=1.5]{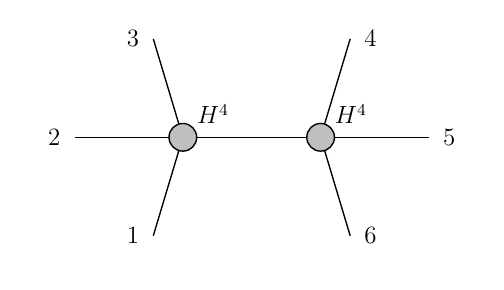}
\caption{Factorization of a six-point amplitude though a pair of $H^4$ vertices.}
\label{644fac}
\end{figure}

The statement that the coefficient functions~$f_6(\Phi^I)$ of the six-derivative terms are quadratically related to the coefficient functions~$f_4(\Phi^I)$ that occur at four-derivative order can also be understood through factorization. First, note that there is no four-point, six-derivative supervertex in a theory of a single Abelian tensor multiplet. This is because such a supervertex must be proportional to $\delta^8(Q)(s+t+u)$, but $s+t+u=0$ in a massless four-point amplitude. There is also no local six-point, six-derivative supervertex that is an~$\mathfrak{so}(5)_R$ singlet. Therefore, the six-point coupling~$H^6$ is part of a non-local superamplitude that is completely determined by its factorization through a pair of four-point supervertices of the form $\delta^8(Q)$, and hence it must be proportional to~$f_4^2$. This factorization channel is shown in Figure~\ref{644fac}.  As in the discussion around~\eqref{absq} and~\eqref{kbsq}, the coefficients of proportionality between~$f_6$ and~$f_4^2$ are fixed by supersymmetry and can be determined by examining any non-trivial set of superamplitudes that obeys the supersymmetry Ward identities. 

\begin{figure}[htb]
\centering
\includegraphics[scale=1.5]{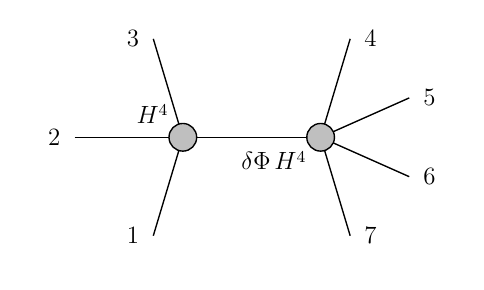}
\caption{Factorization of a $7$-point amplitude though a pair of 4-derivative vertices.}
\label{facwz}
\end{figure}

Finally, we would like to examine the quadratic relation~\eqref{kbsq} between the coefficient~$\Delta k$ of the WZ term in~\eqref{kenwzw} and the coefficient~$b$ of the four-derivative terms~$H^4$. There is a unique five-point supervertex at four-derivative order, which contains the coupling~$\delta\Phi^I H^4$ and arises by expanding the coefficient function in~$f_4(\Phi^I) H^4$. As explained in section 2.1, the WZ term leads to a six-derivative vertex involving seven scalars. The absence of an~$\mathfrak{so}(5)_R$ singlet supervertex at this derivative order implies that this seven-scalar vertex is part of a non-local seven-point superamplitude, which is completely determined by factorization. The only possible factorization channel is displayed in Figure~\ref{facwz}. It involves the five-point supervertex containing~$\delta\Phi^I H^4$ and the four-point supervertex~$\delta^8(Q)$ containing~$H^4$. This establishes the quadratic relation~$\Delta k \sim b^2$ in~\eqref{kbsq}.

\section{Compactification to Five Dimensions}

In this section we consider~$(2,0)$ superconformal theories~$\CT_{\frak g}$ on~$\R^{4,1} \times S^1_R$, where~$S^1_R$ is a spatial circle with radius~$R$ and periodic boundary conditions for the Fermions.\footnote{~As was stated in the introduction, we do not consider outer automorphism twists around the circle.} By the assumption stated in section 1.3, the five-dimensional description, which is valid at energies far below the KK scale~$1 \over R$, is an~$\CN=2$ Yang-Mills theory with gauge algebra~$\frak g$ and gauge coupling~$g^2 \sim R$. We will first describe this Yang-Mills theory, including possible higher-derivative operators that are expected to be present at finite radius~$R$. As in six dimensions, we then explore the Coulomb branch of~$\CT_{\frak g}$ on~$\R^{4,1} \times S^1_R$ using non-renormalization theorems. These allow us to interpolate between the five-dimensional Yang-Mills description, which is valid near the origin, and the effective theory far out on the Coulomb branch, which is simply related to the effective action on the six-dimensional tensor branch that was discussed in section~2. In one direction, the five-dimensional Yang-Mills description leads to information about the six-dimensional theory: the structure of its moduli space, the spectrum of dynamical W-strings on the tensor branch, and the coefficient~$b$ in~\eqref{fourd} and~\eqref{fsolve}, which governs the tensor-branch effective action through six-derivative order. Conversely, the parent theory~$\CT_{\frak g}$ constrains the five-dimensional effective theory: the properties of W-strings in six dimensions restrict~$\frak g$ to be of ADE type, and the conformal invariance of~$\CT_{\frak g}$ requires the leading higher-derivative operators at the origin of the five-dimensional Coulomb branch to be absent. We also show that the only~$(2,0)$ theories with~$\frak g = \frak u(1)$ are free Abelian tensor multiplets. Some of the arguments in this section are standard, while others offer a different point of view on known results (see for instance~\cite{Seiberg:1997ax, Witten:2009at,Gaiotto:2010be, moorefklect}). We include them here partly to render the discussion self-contained, and partly to emphasize that the conclusions only rely on the assumption stated in section~1.3.

\subsection{The Yang-Mills Description at the Origin}

The five-dimensional Lagrangian~$\SL^{(5)}_0$ at the origin of the Coulomb branch is a weakly-coupled Yang-Mills theory with gauge algebra~$\frak g$. For now, we will take~$\frak g$ to be the compact real form of a simple Lie algebra; the case~$\frak g = \frak u(1)$ is discussed in section 3.4. More properly, we should choose a gauge group~$G$, whose Lie algebra is~$\frak g$. This choice arises, because the six-dimensional theory~$\CT_{\frak g}$ generally does not possess a conventional partition function, but rather a family of partition functions valued in a finite-dimensional vector space -- sometimes referred to as the space of conformal blocks~\cite{Witten:2009at}.\footnote{~In this sense, many~$(2,0)$ theories~$\CT_{\frak g}$ require a slight generalization of standard quantum field theory~\cite{Witten:2009at}. It is always possible to obtain theories with a standard partition function by appropriately adding free decoupled tensor multiplets~\cite{Gaiotto:2014kfa}.} The ability to specify the gauge group~$G$ in five dimensions reflects the freedom to choose a partition function in the space of conformal blocks. The ambiguity of the partition function also implies that~$\CT_{\frak g}$ does not respect standard modular invariance, e.g.~when it is compactified on tori~\cite{Witten:2009at}. Much of the discussion below only depends on the Lie algebra~$\frak g$, but we will occasionally encounter global issues.  

We will work in conventions where~$\frak g$ is represented by Hermitian matrices, so that the structure constants are purely imaginary. Since the gauge field~$A = A_\mu dx^\mu$ and the field strength~$F = \half F_{\mu\nu} dx^\mu \wedge dx^\nu = dA -i A \wedge A$ are are valued in~$\frak g$, they are also Hermitian. The same is true for the other fields in the~$\CN=2$ Yang-Mills multiplet: the scalars~$\phi^I$ in the~$\bf 5$ of the~$\frak{so}(5)_R$ symmetry, which is preserved by the circle compactification, and symplectic Majorana Fermions transforming in the fundamental spinor representations of the Lorentz and~$R$-symmetry. As in four dimensions, $A$ and~$\phi^I$ have mass dimension one, while the Fermions have dimension~$3 \over 2$. 

In order to give meaning to the gauge coupling~$g^2$, we must specify a normalization for the gauge field~$A$, and hence the Lie algebra~$\frak g$.\footnote{~We will recall aspects of Lie algebras and Lie groups as they arise in our discussion. For a systematic review in the context of gauge theories, see for instance\cite{Kapustin:2005py,Gaiotto:2010be}} The Lie algebra~$\frak g$ decomposes into a Cartan subalgebra~$\frak t_{\frak g}$ and root vectors~$e_\alpha$, which diagonalize the adjoint action of the Cartan subalgebra, i.e.~for every~$h \in \frak t_{\frak g}$ and every root vector~$e_\alpha$, 
\begin{equation}
[h, e_\alpha] = \alpha(h) e_\alpha~.
\end{equation}
The real functional~$\alpha \in \frak t_{\frak g}^*$ is the root associated with~$e_\alpha$, and  the set of all roots comprises the root system~$\Delta_{\frak g} \subset \frak t_{\frak g}^*$ of the Lie algebra~$\frak g$. For every root~$\alpha \in \Delta_{\frak g}$, there is a unique coroot~$h_\alpha \in \frak t_{\frak g}$, which together with~$e_{\pm \alpha}$ satisfies the commutation relations of~$\frak{su}(2)$,\footnote{~In these conventions, the eigenvalues of~$h_\alpha$ are always integers, rather than half-integers. }
\begin{equation}\label{sucomm}
[e_\alpha, e_{-\alpha}] = h_\alpha~, \qquad [h_\alpha, e_{\pm \alpha}] = \pm 2 e_{\pm \alpha}~.
\end{equation}
We define a normalized, positive-definite trace~$\Tr_{\frak g}$,
\begin{equation}\label{trdef}
\Tr_{\frak g} = {1 \over 2 h^\vee_{\frak g}} \Tr_{\text{adj}}~,
\end{equation}
where~$h^\vee_{\frak g}$ is the dual Coxeter number. This induces a positive-definite metric~$\langle \cdot, \cdot\rangle_{\frak g}$ on the Cartan subalgebra,
\begin{equation}
\langle h, h'\rangle_{\frak g} \equiv \Tr_{\frak g} (h h')~, \qquad  h, h' \in \frak t_{\frak g}~,
\end{equation} 
and hence also its dual~$\frak t^*_{\frak g}$, which contains the root system~$\Delta_{\frak g}$. The definition in~\eqref{trdef} is in accord with the standard convention that a short co-root~$h_\alpha$, and the corresponding long root~$\alpha$, both satisfy~$\langle h_\alpha, h_\alpha\rangle_{\frak g} = \langle \alpha, \alpha\rangle_{\frak g}= 2$.\footnote{~As an example, consider~$\frak g = \frak{su}(2)$, where~$h^\vee_{\frak{su}(2)} = 2$. The commutation relations~$[e_+, e_-] = h$ and~$[h, e_\pm] = \pm 2 e_\pm$ imply that~$h_{\text{adj}} = \text{diag}(2,0,-2)$. Therefore~$\Tr_{\text{adj}}(h^2) = 8$ and~$\Tr_{\frak{su}(2)} \left(h^2\right) = 2$.
}  
In these conventions, the instanton number on~$S^4$, which can take all possible integer values, is given by the following expression~\cite{Bernard:1977nr},
\begin{equation}
{1 \over 8 \pi^2} \int_{S^{4}} \mathrm{Tr}_{\mathfrak{g}}(F\wedge F) \in \Z~. \label{minimalinst}
\end{equation}

The two-derivative terms in the five-dimensional low-energy theory are given by the usual Yang-Mills Lagrangian,
\begin{align}
\SL_0^{(5)}=& -\frac{1}{2g^{2}}\mathrm{Tr}_{\mathfrak{g}}\left(F\wedge * F +\sum_{I = 1}^ 5 D_{\mu}\phi^{I}D^{\mu}\phi^{I} - {1 \over 8} \sum_{I, J = 1}^5 \left[\phi^I, \phi^J\right]^2 \right) \cr
& +\left(\text{Fermions}\right) + \left(\text{higher-derivative terms}\right)~, \label{5dkin}
\end{align}
where~$D = d - i [A, \cdot]$ is the covariant derivative in the adjoint representation. In five dimensions the gauge coupling~$g^2$ has dimensions of length. It follows from the scale-invariance of the six-dimensional theory that~$g^2$ must be proportional to the compactification radius~$R$. However, our assumptions do not obviously fix the constant of proportionality. We can motivate the answer by appealing to the following intuitive picture, which can be made precise in string constructions: at the origin of the Coulomb branch, $\CN=2$ Yang-Mills theories in five dimensions admit particle-like solitons, which are the uplift of four-dimensional instantons. Their mass is proportional to~$n \over g^2$, where~$n$ is the instanton number, and since~$g^2 \sim R$ it is tempting to interpret them as massive KK modes of the six-dimensional theory. We can fix~$g^2$ in terms of~$R$ by demanding that the mass of the minimal~$\frak g$-instanton-soliton in flat space (more precisely on~$S^4$) coincides with the minimal KK mass~${1 \over R}$. This leads to
\begin{equation}
g^{2}=4 \pi^2 R~. \label{instanton}
\end{equation}
Note that this picture involves particles whose mass is necessarily of the same order as the cutoff of the five-dimensional effective theory. Below we will show that~\eqref{instanton} can be reliably derived in effective field theory, by extrapolating between five and six dimensions along the Coulomb branch.

The higher-derivative terms in~\eqref{5dkin} are constrained by~$\CN=2$ supersymmetry and~$\frak{so}(5)_R$ symmetry. These constraints were analyzed in \cite{Movshev:2009ba,Bossard:2010pk,Chang:2014kma}, with the following conclusions:

\begin{itemize} 
\item The leading irrelevant operators that can appear in~\eqref{5dkin} occur at four-derivative order, as supersymmetric completions of non-Abelian~$F^4$ terms. There are two independent such terms -- a single-trace operator and a double-trace operator.\footnote{~When~$\frak{g}=\frak{su}(2)$, there is only one independent operator, due to trace relations.} Schematically,
\begin{subequations}\label{F4}
\begin{align}
& x\,g^{6}~ t^{\mu_{1}\nu_{1}\mu_{2}\nu_{2}\mu_{3}\nu_{3}\mu_{4}\nu_{4}}~\mathrm{Tr}_{\mathfrak{g}}\left(F_{\mu_{1}\nu_{1}}F_{\mu_{2}\nu_{2}}F_{\mu_{3}\nu_{3}}F_{\mu_{4}\nu_{4}}\right)+\cdots \subset \SL_0^{(5)}~, \\
&y\,g^{6}~t^{\mu_{1}\nu_{1}\mu_{2}\nu_{2}\mu_{3}\nu_{3}\mu_{4}\nu_{4}}~\mathrm{Tr}_{\mathfrak{g}}\left(F_{\mu_{1}\nu_{1}}F_{\mu_{2}\nu_{2}}\right) \, \mathrm{Tr}_{\mathfrak{g}}\left(F_{\mu_{3}\nu_{3}}F_{\mu_{4}\nu_{4}}\right)+\cdots \subset \SL_0^{(5)}~,
\end{align}
\end{subequations}
where the ellipses denote the supersymmetric completions involving scalars and Fermions, and the powers of the cutoff~$g^2$ are fixed so that~$x, y$ are dimensionless constants. The tensor~$t^{\mu_{1}\nu_{1}\mu_{2}\nu_{2}\mu_{3}\nu_{3}\mu_{4}\nu_{4}}$, which determines how the spacetime indices are contracted, is constructed out of the metric~$\eta_{\mu\nu}$.\footnote{~The tensor~$t^{\mu_{1}\nu_{1}\mu_{2}\nu_{2}\mu_{3}\nu_{3}\mu_{4}\nu_{4}}$ occurs for maximally supersymmetric Yang-Mills theories in all dimensions. For instance, it is discussed in chapter~12 of~\cite{Polchinski:1998rr}.} One particular linear combination of~$x$ and~$y$ appears in the non-Abelian DBI Lagrangian describing coincident~D4-branes~\cite{Bergshoeff:2001dc}. 

Both operators in~\eqref{F4} are~$\half$-BPS -- they can be written as~$Q^8$ acting on a gauge-invariant local operator constructed out of fields in the Yang-Mills multiplet -- and they are the only such operators that preserve the~$\frak{so}(5)_R$ symmetry. (See~\cite{Chang:2014kma} for a discussion of~$\half$-BPS operators that break the~$R$-symmetry.) On the Coulomb branch, where~$F$ is restricted to the Cartan subalgebra, these operators give rise to~$\half$-BPS, four-derivative terms that can be tracked along the moduli space. Below we will use this to argue that~$x$ and~$y$ must vanish for all~$(2,0)$ SCFTs compactified on~$S^1_R$. 

\item At six-derivative order, there is a single~${1\over 4}$-BPS, double-trace operator of the schematic form~$\Tr_{\frak g}^2  D^2 F^4$ \cite{Bossard:2010pk}. Generically, this operator induces six-derivative terms on the Coulomb branch. However, we will see below that it does not contribute on Coulomb branches described by a single Abelian vector multiplet.

\item Starting at six-derivative order, there are full~$D$-terms, which can be written as~$Q^{16} \CO$, where~$\CO$ is a gauge-invariant local operator. The leading such~$\CO$ is the Konishi-like single-trace operator~$\sum_{I = 1}^5 \Tr_{\frak g}\left(\phi^I \phi^I\right)$, and the corresponding~$D$-term is schematically given by~$ {\rm Tr}_{\mathfrak{g}} D^2 F^4$. This term is believed to be present in~$\SL_0^{(5)}$, since it is needed to absorb a six-loop divergence of the two-derivative Yang-Mills theory~\cite{Bern:2012di}. Below, we will show that full~$D$-terms can only affect the Coulomb-branch effective action at eight-derivative order or higher.

\end{itemize}
Having described the theory at the origin, we will now explore its Coulomb branch.

\subsection{Two-Derivative Terms and BPS States on the Coulomb Branch}

The scalar potential in~\eqref{5dkin} restricts the adjoint-valued scalars to a Cartan subalgebra~$\frak t_{\frak g} \subset \frak g$. Therefore, the Coulomb branch is parametrized by~$\langle \phi^I \rangle \in \R^5 \otimes \frak t_{\frak g} / \CW_{\frak g}$, where~$\R^5$ transforms in the~$\bf 5$ of the~$\frak{so}(5)_R$ symmetry and~$\CW_{\frak g}$ is the Weyl group of~$\frak g$, which acts on~$\frak t_{\frak g}$. At a generic point on the Coulomb branch, the low-energy theory consists of~$r_{\frak g}$ Abelian vector multiplets, with scalars~$\varphi_i^I$ and field-strengths~$f_i$, which are permuted by the Weyl group. Their embedding into the non-Abelian fields at the origin can be written as follows,
\begin{equation}\label{hexp}
\phi^I = \sum_{i =1}^{r_{\frak g}} \, h_i \, \varphi_i^I ~, \qquad F = \sum_{i = 1}^{r_{\frak g}} \, h_i   f_i ~.
\end{equation}
Here we use a basis of simple coroots~$h_i$ for the Cartan subalgebra, which are associated with the~$r_{\frak g}$ simple roots~$\alpha_i$ via~\eqref{sucomm}. Their commutation relations with the root vectors~$e_{\pm i} = e_{\pm \alpha_i}$ are determined by the Cartan matrix~$C_{ij}$,
\begin{equation}
[h_i, h_j] = 0~, \qquad [e_{+i}, e_{-j}] = \delta_{ij} h_j~, \qquad [h_i, e_{\pm j}] = \pm C_{ji} e_{\pm j}~. \label{algconv}
\end{equation}
In these equations, the repeated index~$j$ is not summed. Substituting~\eqref{hexp} into~\eqref{5dkin}, we obtain the leading two-derivative effective action on the Coulomb branch,
\begin{equation}\label{freeu1}
\SL^{(5)}_{\text{Coulomb}} = -{1 \over 2 g^2} \Omega_{ij} \left(f_i \wedge * f_j + \sum_{I =1}^5 \d_\mu \varphi^I_i \d^\mu \varphi_j^I \right) + \left(\text{Fermions}\right) + \cdots~,
\end{equation}
where the ellipsis denotes a variety of possible corrections that will be discussed below. The kinetic terms are determined by a symmetric, positive-definite matrix,
\begin{equation}\label{omegadef}
\Omega_{ij} = \Tr_{\frak g} \left(h_i h_j\right) = \langle h_i, h_j\rangle_{\frak g}~.
\end{equation}
Note that the normalization of the Abelian gauge fields~$f_i$ is meaningful, since they are embedded in the non-Abelian~$F$ according to~\eqref{hexp}. More precisely, the fluxes of the~$f_i$ are quantized in units dictated by the gauge group~$G$, i.e.~they depend on the global properties of~$G$, not just on its Lie algebra. This will not affect the present discussion, but it will play a role when we discuss the compactification to four dimensions in section~5. 

We obtained~\eqref{freeu1} classically, by restricting the fields in~\eqref{5dkin} to the Cartan subalgebra. There are two possible kinds of corrections: quantum corrections that modify the two-derivative terms in~\eqref{freeu1}, and higher-derivative corrections. The latter are present and will be discussed in section~3.3. The former are known be absent for maximally supersymmetric Yang-Mills theories in all dimensions.\footnote{~Just as in the discussion around~\eqref{kinnrt}, this can be shown by expanding any moduli-dependent two-derivative terms around a fixed vev~$\langle \varphi_i^I\rangle$ and noting that the free two-derivative theory does not admit supersymmetric deformations containing three fields and two derivatives~\cite{Bossard:2010pk,Chang:2014kma}.} Therefore the geometry of the Coulomb branch is dictated by the classical theory,
\begin{equation}\label{5d6dm}
\CM_{\frak g} = \R^5 \otimes \frak t_{\frak g} / \CW_{\frak g}~,
\end{equation}
with the flat metric~\eqref{omegadef}. The only singularities are of orbifold type and occur at the boundaries of~$\CM_{\frak g}$, where part of the gauge symmetry is restored. The allowed patterns of gauge symmetry breaking and restoration are governed by adjoint Higgsing. Therefore, the breaking pattern~$\frak g \rightarrow \frak h \oplus \frak u(1)^n$ with~$\frak h$ semisimple and~$n \leq r_{\frak g}$ is allowed if the Dynkin diagram of~$\frak h$ can be obtained from the Dynkin diagram of~$\frak g$ by deleting~$n$ nodes (see for instance~\cite{Slansky:1981yr}). 

Since the geometry of the Coulomb branch is rigid, it can be extrapolated to vevs~$|\langle \phi_i^I \rangle|$ that are much larger than the KK scale~$1 \over R$, i.e.~there are no corrections due to KK modes. Therefore, the moduli spaces in five and six dimensions are identical -- they are both given by~\eqref{5d6dm} -- and the two-derivative effective actions that describe them are simply related. Explicitly, the fields in five dimensions are obtained by reducing the six-dimensional fields to zero modes along~$S^1_R$,
\begin{equation}\label{5d6dfields}
\Phi^I_i \rightarrow {1 \over  2 \pi R}\varphi^I_i ~, \qquad H_i \rightarrow {1 \over 2 \pi R} \left(f_i \wedge dx^5 + *^{(5)} f_i\right)~.
\end{equation} 
Here~$x^5 \sim x^5 + 2 \pi R$ parametrizes the circle and~$*^{(5)}$ denotes the five-dimensional Hodge star, so that~$H_i = *H_i$ in six dimensions. Note that the units of quantization  for the fluxes of~$H_i$ are dictated by those of~$f_i$, which are in turn determined by the five-dimensional gauge group~$G$. Using~\eqref{5d6dfields}, the two-derivative action~\eqref{freeu1} can be uplifted to six dimensions,
\begin{equation}\label{6dlift}
-{\pi R \over  g^2} \, \Omega_{ij} \Big(H_i \wedge * H_j + \sum_{I =1}^5 \d_\mu \Phi^I_i \d^\mu \Phi_j^I \Big) + \left(\text{Fermions}\right)  \subset \SL_{\text{tensor}}~.
\end{equation} 

As in section~2, the quadratic Lagrangian for the self-dual fields~$H_i$ in~\eqref{6dlift} vanishes, since~$\Omega_{ij}$ symmetric while~$H_i \wedge H_j$ is antisymmetric, and hence their non-canonical kinetic terms may seem meaningless. This is not the case, in part because the fluxes of the~$H_i$ are quantized in definite units. Therefore, global observables, such as partition functions on closed manifolds, are sensitive to~$\Omega_{ij}$ (see~\cite{Witten:2007ct} and references therein). Here we will use the fact that the kinetic terms in~\eqref{6dlift} determine the six-dimensional Dirac pairing for string sources that couple to the~$H_i$. For a string with worldsheet~$\Sigma_2$ and charges~$q_i$, 
\begin{equation}\label{6dsc}
dH_i = q_i \delta_{\Sigma_2} \qquad \Longleftrightarrow \qquad q_i = \int_{\Sigma_3} H_i~,
\end{equation} 
where~$\delta_{\Sigma_2}$ is a unit delta function localized on~$\Sigma_2$, which is linked by the three-cycle~$\Sigma_3$. The integer-valued Dirac pairing between two strings with charges~$q_i, q'_i$ is then given by~\cite{Deser:1997mz},\footnote{~If the~$H_i$ were not self dual, the pairing in~\eqref{6diracpair} would be valued in~$\half \Z$ (as in four-dimensional electrodynamics) rather than in~$\Z$. The relative factor of~$\half$ arises because the self-dual and the anti-self-dual parts of~$H_i$ contribute equally to the angular momentum, whose quantization leads to the Dirac condition.}
\begin{equation}\label{6diracpair}
{R \over g^2} \, \Omega_{ij} q_i q'_j \in \Z~.
\end{equation}
This gives operational meaning to the non-trivial kinetic terms for the~$H_i$ in~\eqref{6dlift}. The importance of such terms was recently emphasized in~\cite{Intriligator:2014eaa}, and also played a role in~\cite{Ohmori:2014kda}.  

A set of candidate string sources for the~$H_i$ is furnished by the dynamical W-strings on the tensor branch, which were mentioned at the end of section~2.1. When the theory is compactified on a circle, it is natural to compare them to the~BPS states of the five-dimensional Yang-Mills theory on the Coulomb branch (see~\cite{Gaiotto:2010be,Tachikawa:2011ch} for a brief summary). We will focus on the electrically charged W-Bosons, which correspond to roots~$\alpha \in \Delta_{\frak g}$, and magnetically charged monopole strings, which correspond to coroots~$h_\alpha$. Both the~W-Bosons and the monopole strings are~$\half$-BPS and their masses are proportional to the vevs~$\langle \varphi^I_i\rangle$. As such, they become parametrically light near the origin of the Coulomb branch, where their properties are completely determined by the low-energy Yang-Mills theory. On the other hand, it is believed that such~$\half$-BPS states can be reliably extrapolated to large vevs, where the theory is effectively six-dimensional. In that regime, both the W-Bosons and the monopole strings must arise from six-dimensional W-strings. The former correspond to strings that wrap~$S^1_R$, as in~\cite{Witten:1995zh}, while the latter describe strings that lie in the five non-compact dimensions. Therefore, the electric charges of the W-Bosons and the magnetic charges of the monopoles must arise from the same set of W-string charges in six dimensions. 

This six-dimensional requirement leads to constraints on the five-dimensional effective theory. Consider the W-Boson corresponding to a fixed simple root~$\alpha_i$. It follows from~\eqref{algconv} that its electric charges~$(e_i)_j$ with respect to the Abelian gauge fields~$f_j$ are given by the entries~$C_{ij}$ in the~$i^{\text{th}}$ row of the Cartan matrix. These charges can be measured by evaluating the~$j^{\text{th}}$ electric flux across a Gaussian surface~$\Sigma^i_3$ that surrounds the W-Boson, 
\begin{equation}\label{elch}
(e_i)_j = C_{ij} = {\Omega_{jk} \over g^2} \int_{\Sigma^i_3}  * f_k~.
\end{equation}
Similarly, the magnetic charges~$(m_i)_j$ of the monopole-string corresponding to the simple coroot~$h_i$, measured with respect to~$f_j$, are given by
\begin{equation}\label{magch}
(m_i)_j = \delta_{ij} = {1 \over 2 \pi} \int_{\Sigma^i_2} f_j~,
\end{equation}
where~$\Sigma^i_2$ links the monopole string. If we use~\eqref{5d6dfields} to express the integrals in~\eqref{elch} and~\eqref{magch} as integrals of the six-dimensional three-form flux~$H_i$ over~$\Sigma_3^i$ and~$\Sigma_2^i \times S_R^1$, and we demand that these integrals measure the same six-dimensional W-string charge~$(q_i)_j$, as defined in~\eqref{6dsc}, then we obtain
\begin{equation}\label{result}
(q_i)_j = 2 \pi \delta_{ij}~, \qquad C_{ij} = {4 \pi^2 R \over g^2} \,\Omega_{ij}~.
\end{equation}
Since~\eqref{omegadef} implies that~$\Omega_{ij}$ is symmetric, the same must be true for the Cartan matrix~$C_{ij}$. However, this is only possible if~$\frak g$ is simply laced. Note that this argument crucially relies on properties of the six-dimensional parent theory, specifically its W-strings. It does not imply that all five-dimensional~$\CN=2$ Yang-Mills theories with non-ADE gauge groups are inconsistent. For instance, such theories arise by activating outer automorphism twists around the compactification circle, as in~\cite{Vafa:1997mh,Witten:2009at,Tachikawa:2011ch}. However, this ruins the symmetry between wrapped and unwrapped W-strings that lead to~\eqref{result}. Previous arguments for the ADE restriction used anomaly cancellation on the W-string worldsheet~\cite{Henningson:2004dh}, or self-duality and modular invariance in~$(2,0)$ theories with standard partition functions~\cite{Seiberg:1997ax,Seiberg:2011dr}. In section~4.3 we will describe another argument for the ADE restriction that only relies on the consistency of the low-energy effective theory on the six-dimensional tensor branch. 

We can use~\eqref{result} to derive the relationship between~$g^2$ and~$R$. In general, the Cartan matrix is given by 
\begin{equation}
C_{ij} = {2 \langle \alpha_i, \alpha_j\rangle_{\frak g} \over  \langle \alpha_j, \alpha_j\rangle_{\frak g}}~.
\end{equation}
If~$\frak g$ is simply laced, then all~$\alpha_j$ have the same length; in our conventions~$ \langle \alpha_j, \alpha_j\rangle_{\frak g} = 2$. Therefore~$C_{ij}$ precisely coincides with~$\Omega_{ij}$ as defined in~\eqref{omegadef}, so that~\eqref{result} reduces to
\begin{equation}
g^{2}=4 \pi^2 R~, \label{gaugeagain}
\end{equation}
in agreement with~\eqref{instanton}. Together with~\eqref{result}, this shows that the Dirac quantization condition~\eqref{6diracpair} for the W-strings amounts to the statement that the entries of the Cartan matrix are integers.  

\subsection{Four-Derivative Terms on the Coulomb Branch}

The higher-derivative terms in~$\SL_{\text{Coulomb}}^{(5)}$ can arise in two ways: classically, by restricting higher-derivative terms that are already present in the effective Lagrangian~$\SL^{(5)}_0$ at the origin to the Coulomb branch; and quantum mechanically, by integrating out W-Bosons. In general, understanding these corrections ultimately requires detailed knowledge of the higher-derivative terms in~$\SL_0^{(5)}$, including~$D$-terms. However, just as in six dimensions, the first few orders in the derivative expansion of~$\SL_{\text{Coulomb}}^{(5)}$ are protected by non-renormalization theorems. 

We again restrict the discussion to rank one Coulomb branches with a single Abelian vector multiplet that arise by breaking~$\frak g\rightarrow \frak h \oplus \frak u(1)$. The vevs~$\langle \varphi^I\rangle$ of the scalars in this vector multiplet still break the~$\frak{so}(5)_R$ symmetry to~$\frak{so}(4)_R$, which leads to four NG Bosons, but since the five-dimensional theory is not conformal, there is no dilaton. Nevertheless, it is useful to introduce the radial variable
\begin{equation}
\psi = \left(\sum_{I = 1}^5 \varphi^I \varphi^I \right)^\half~.
\end{equation}
As in section~3.2, the kinetic terms in~$\SL^{(5)}_{\text{Coulomb}}$ are determined by the embedding of~$\varphi^I, f$ into the non-Abelian fields~$\phi^I, F$ at the origin,
\begin{equation}\label{nonabeemb}
\phi^I = t \varphi^I~, \qquad F = t f~, \qquad t \in \frak t_{\frak g}~.
\end{equation}
Here~$t \in \frak t_{\frak g}$ is a Cartan generator whose commutant in~$\frak g$ is~$\frak h \oplus \frak u(1)$,\footnote{~We denote the Cartan generator by~$t$ rather than~$h$, in order to avoid confusion with the subalgebra~$\frak h \subset \frak g$.} so that the kinetic terms on the Coulomb branch are given by
\begin{equation}\label{kt}
-{1 \over 2 g^2} \Tr_{\frak g}\left(t^2\right) \left(f \wedge * f + \sum_{i = I}^5 \d_\mu \varphi^I \d^\mu \varphi^I\right) + \left(\text{Fermions}\right) \; \subset \; \SL^{(5)}_{\text{Coulomb}}~.
\end{equation}

As in maximally supersymmetric Yang-Mills theories in other dimensions~\cite{Paban:1998ea,Paban:1998mp,Paban:1998qy,Sethi:1999qv}, the dependence of the first several higher-derivative terms in~$\SL^{(5)}_{\text{Coulomb}}$ on the scalars~$\varphi^I$ is tightly constrained. Here we will follow the logic of section~2.2: first expand the moduli-dependent coefficient functions in the effective Lagrangian around a fixed vev, and then impose the constraints of supersymmetry on the resulting local operators. It follows from the analysis in~\cite{Movshev:2009ba,Bossard:2010pk,Chang:2014kma} that the possible supersymmetric deformations of single free Abelian vector multiplet take exactly the same form as the~$F$- and~$D$-term deformations~\eqref{fterm} and~\eqref{dterm} of a free Abelian tensor multiplet in six dimensions. The former are four-derivative terms of the form~$Q^8 \left(\varphi^{(I_1} \cdots \varphi^{I_n)} - \left(\text{traces}\right)\right)$ with~$n \geq 4$, which transform as symmetric, traceless~$(n-4)$-tensors of~$\frak{so}(5)_R$. The latter first arise at eight-derivative order, and there are no independent six-derivative deformations. Therefore the conclusions of section~2.2 still apply, with minimal modifications. In particular, the four-derivative terms in~$\SL^{(5)}_{\text{Coulomb}}$ are now controlled by two dimensionless coefficients~$b^{(5)}$ and~$c^{(5)}$, which we define as follows,
\begin{equation}\label{b5def}
\left({b^{(5)} \over \psi^3} + c^{(5)} g^6\right) \left(\d\psi\right)^4 \subset \SL^{(5)}_{\text{Coulomb}}~. 
\end{equation} 
As in section~2.2, the~$\psi$-dependence of the coefficient function in parentheses follows from the fact it must be harmonic and~$\frak{so}(5)_R$ invariant. Since the theory is not conformal, the term proportional to~$c^{(5)}$ is not a priori forbidden. As in six dimensions, all six-derivative terms in~$\SL^{(5)}_{\text{Coulomb}}$ are determined by the lower-order terms. Here we will focus on the four-derivative terms, and in particular on~\eqref{b5def}. As in section~2.3, these conclusions also follow from considerations involving superamplitudes (see also appendix A). 

Since the dependence of~\eqref{b5def} on the dimensionful gauge coupling~$g$ is completely determined by the non-renormalization theorem, we can fix the coefficients~$b^{(5)}$ and~$c^{(5)}$ at parametrically weak coupling. In this limit, $c^{(5)}$ can only arise from a classical contribution due to four-derivative terms in the Lagrangian~$\SL_0^{(5)}$ at the origin. The only such terms are the two~$\half$-BPS terms described around~\eqref{F4}, with independent dimensionless coefficients~$x, y$. Restricting the non-Abelian gauge fields at the origin to the Cartan direction~$t$ shows that
\begin{equation}\label{chh}
c^{(5)} \sim x \, \mathrm{Tr}_{\mathfrak{g}}(t^{4})+ \, y\left(\mathrm{Tr}_{\mathfrak{g}}(t^{2})\right)^2 ~.
\end{equation}
We can change~$t$ by considering different rank one adjoint breaking patterns, which samples different linear combinations of~$x$ and~$y$.\footnote{~The only exception is~$\frak g = \frak{su}(2)$, but in that case~$x$ and~$y$ are linearly dependent due to trace relations.} 

The fixed dependence of~\eqref{b5def} on~$\psi$ enables us to extrapolate to large vevs~$|\langle \psi\rangle|$ and compare with the six-dimensional effective Lagrangian~$\SL_{\text{tensor}}$, as we did for the kinetic terms in section~3.2. However, in that regime, the scale-invariance of the six-dimensional theory forbids the constant~$c^{(5)}$ in~\eqref{b5def}. By comparing with~\eqref{chh} for different choices of~$t$, we conclude that both~$x$ and~$y$ must vanish. Therefore, the leading possible higher-derivative terms~\eqref{F4} at the origin of the five-dimensional Coulomb branch are absent. For~$\frak g = \frak{su}(2)$, this was argued in~\cite{Lin:2015zea} via a comparison with little string theory. Here we see that it is a simple and general consequence of the fact that the six-dimensional~$(2,0)$ theory is scale invariant. 

\bigskip
\bigskip

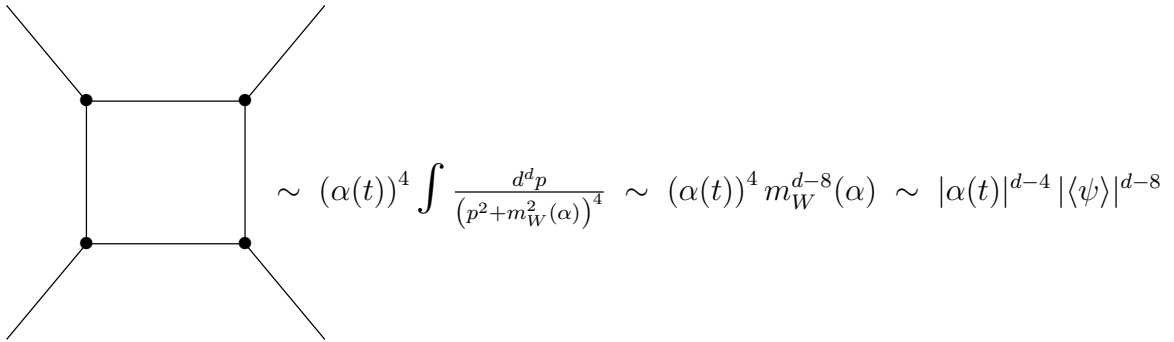
\begin{figure}[h]

\qquad\xymatrix  @R=1pc {
*=0{\phantom{\bullet}}\ar@{-}[ddr]&&&&   *=0{\phantom{\bullet}}\ar@{-}[ddl]&&&&\\
\\
&*=0{\bullet} \ar@{-}[rr] \ar@{-}[ddd]&&   *=0{\bullet}\ar@{-}[ddd]&&&&&\\
 \\
 &&&&&&&&*=0{ \qquad \qquad \quad\sim \; \left(\alpha(t)\right)^{4}\bigintsss {d^{d}p \over \left(p^{2}+m^2_W(\alpha)\right)^4} \; \sim \; \left(\alpha(t)\right)^{4}m^{d-8}_{W}(\alpha)\; \sim \; |\alpha(t)|^{d-4} \, |\langle \psi\rangle|^{d-8}}
 \\
& *=0{\bullet} \ar@{-}[rr]&&   *=0{\bullet}&&&&&\\
\\
*=0{\phantom{\bullet}}\ar@{-}[uur]&&&&   *=0{\phantom{\bullet}}\ar@{-}[uul]&&&&}  

 \caption{One-loop box diagram in~$d$ spacetime dimensions. The external lines are associated with the Cartan generator~$t$ and the W-Bosons running in the loop correspond to a root~$\alpha$.}
  \label{loopfig}
 \end{figure}
 
 \medskip

We now apply the same logic to the coefficient~$b^{(5)}$ in~\eqref{b5def}. Since it does not depend on the gauge coupling~$g$, it can only arise from the two-derivative Yang-Mills Lagrangian in~\eqref{5dkin} by integrating out massive W-Bosons at one loop. Upon turning on a vev along the Cartan element~$t$, the gluons in the adjoint representation of~$\frak g$ decompose into the massless gluons of~$\frak h$ and the~$\frak u(1)$ photon, as well as massive W-Bosons. The later are labeled by roots~$\alpha \in \Delta_{\frak g}$ that do not reside in the root system of~$\frak h$, i.e~$\alpha \in \Delta_{\frak g} \backslash \Delta_{\frak h}$. The~$\frak u(1)$ charge of a W-Boson labeled by~$\alpha$ is given by~$\alpha(t)$. Carrying out the one-loop computation expresses~$b^{(5)}$ as a sum over W-Bosons, weighted by their~$\frak u(1)$ charges,
\begin{equation}\label{b5loop}
b^{(5)} = {1 \over 128 \pi^2} \sum_ {\alpha \in \Delta_{\frak g} \backslash \Delta_{\frak h}} |\alpha(t)|~.
\end{equation}
Up to an overall constant, this result can be understood using a simple scaling argument: the coefficient~$b^{(5)}$ is determined by the one-loop scalar box integral in Figure~\ref{loopfig}, with W-Bosons labeled by roots~$\alpha$ running in the loop (see for instance~\cite{Elvang:2013cua}). It is instructive to examine this integral as a function of the spacetime dimension~$d$. There are four powers of~$|\alpha(t)|$ that arise from the vertices. If the integral over loop momenta is finite, as is the case here, it scales like~$d-8$ powers of the W-Boson mass~$m_W(\alpha)$.  Therefore, the diagram in Figure~\ref{loopfig} is proportional to~$|\alpha(t)|^{d-4} |\langle \psi\rangle|^{d-8}$. For~$d = 5$ this is consistent with~\eqref{b5loop}, since~$b^{(5)}$ multiplies~$\psi^{-3}$ in~\eqref{b5def}. In the special case~$d = 4$, the charges~$|\alpha(t)|$ cancel and the diagram is simply proportional to~$n_W$,  the total number of W-Bosons.\footnote{~This cancellation can be understood in terms of the conformal symmetry of four-dimensional~$\CN=4$ Yang-Mills theory. The four-derivative terms generated by the loop integral in Figure~\ref{loopfig} are proportional to~$\Delta a^{(4)}$, the difference between the four-dimensional $a$-anomaly of the UV and IR theories~\cite{Komargodski:2011vj}. Since the~$a$-anomaly of an~$\CN=4$ theory with gauge algebra~$\frak g$ is proportional to~$d_{\frak g}$, it follows that~$\Delta a^{(4)} \sim n_W$.} Therefore, it is misleading to extrapolate the four-dimensional answer to other dimensions, since it does not correctly capture the sum over charges. We will encounter a similar fallacy in section~5. 

Having determined the coefficient~$b^{(5)}$ in~\eqref{b5def}, we can use the fact that we know the exact~$\psi$-dependence of this term to extrapolate it to large vevs. It can then be compared to the term~\eqref{fourd} in the six-dimensional effective Lagrangian on the tensor branch, which we repeat here for convenience
\begin{equation}\label{fourdii}
b \, {(\d \Psi)^4 \over \Psi^3} \subset\SL_{\text{tensor}}~.
\end{equation}
The coefficient~$b$ was defined in a normalization where the six-dimensional dilaton field~$\Psi$ has canonical kinetic terms. (Note that~\eqref{fourdii} is not invariant under rescalings of~$\Psi$.) We must therefore appropriately renormalize the five-dimensional field~$\psi$ to eliminate the non-canonical kinetic terms in~\eqref{kt}. Also taking into account factors of~$2 \pi R$ that arise in the transition from six to five dimensions (see the discussion around~\eqref{5d6dfields} and~\eqref{6dlift}) gives
\begin{equation}
b = \left({g^2 \over 2 \pi R \Tr_{\frak g}(t^2)} \right)^\half b^{(5)}~.
\end{equation}
Substituting~$g^2 = 4 \pi^2 R$ from~\eqref{gaugeagain} and~$b^{(5)}$ from~\eqref{b5loop} then leads to 
\begin{equation}\label{bfinal}
b = \left({1 \over 8192 \pi^3 \Tr_{\frak g}(t^2)} \right)^\half  \sum_ {\alpha \in \Delta_{\frak g} \backslash \Delta_{\frak h}} |\alpha(t)|~.
\end{equation}
Note that rescaling~$t$ does not change the right-hand side of this formula. 

Having determined the coefficient~$b$ that governs the four-derivative terms on the tensor branch, we can use the relations~\eqref{absq} and~\eqref{kbsq} to determine the coefficients of the six-dimensional WZ terms~\eqref{awzw} and~\eqref{kenwzw}. As was emphasized in~\cite{Intriligator:2000eq}, the~$R$-symmetry WZ term~\eqref{kenwzw} vanishes when it is reduced to zero modes along~$S^1_R$, because it contains a derivative along the circle. Nevertheless, we can determine its coefficient by examining the four-derivative terms, whose reduction to five dimensions is non-trivial.

\subsection{The Abelian Case}

When~$\frak g = \frak u(1)$,  the description at the origin of the five-dimensional Coulomb branch involves a single Abelian~$\CN=2$ vector multiplet, possibly deformed by higher derivative terms. One difference to the non-Abelian case is that the relation between the gauge coupling~$g^2$ and the compactification radius~$R$ is no longer fixed by considerations involving BPS states on the Coulomb branch, as in section~3.2. However, the non-renormalization theorems discussed in section~3.3 still apply. Since the Abelian theory does not give rise to any W-Bosons, the constant~$b$ computed in~\eqref{bfinal} vanishes. As was discussed around~\eqref{fourd}, this can only happen if the six-dimensional theory on the tensor branch is locally free. Therefore, the only~$(2,0)$ SCFTs that give rise to~$\frak u(1)$ gauge theories in five dimensions are locally described by free Abelian tensor multiplets.  

\section{Applications}

In this section we will combine the results of sections~2 and~3 to compute the Weyl anomaly~$a_{\frak g}$ and the~$R$-symmetry anomaly~$k_{\frak g}$ for all~$(2,0)$ SCFTs~$\CT_{\frak g}$. We also prove that~$a_{\frak g}$ strictly decreases under every RG flow that preserves~$(2,0)$ supersymmetry. Finally, we use our computation of~$k_{\frak g}$ to offer another argument for the ADE restriction on the Lie algebra~$\frak g$. 

\subsection{Computing the Anomalies~$a_{\frak g}$ and~$k_{\frak g}$}

Upon breaking~$\frak g\rightarrow \frak h \oplus \frak u(1)$ in six dimensions, the anomaly mismatch between the UV theory~$\CT_{\frak g}$ and the IR theory, which consists of the interacting SCFT~$\CT_{\frak h}$ and an Abelian tensor multiplet, is given by~\eqref{absq} and~\eqref{kbsq}, which we repeat here,
\begin{equation}\label{deltaakii}
\Delta a = a_{\frak g} - \left(a_{\frak h} + 1\right) = \frac{98304 \, \pi^{3}}{7} \,b ^{2}~, \qquad \Delta k = k_{\frak g} - k_{\frak h} = 6144 \pi^3 \, b^2~.
\end{equation}
The constant~$b$ was determined in~\eqref{bfinal} via  a one-loop computation in five dimensions. Substituting~\eqref{bfinal} into~\eqref{deltaakii}, we find
\begin{equation}\label{akx}
\Delta a = {12 \over 7} X ~, \qquad \Delta k = {3\over 4} X~, \qquad X = {1 \over \Tr_{\frak g}(t^2)} \left(\sum_ {\alpha \in \Delta_{\frak g} \backslash \Delta_{\frak h}} |\alpha(t)|\right)^2~.
\end{equation}
Note that the denominator of~$X$ can also be expressed as a sum over~W-Bosons using~\eqref{trdef},
\begin{equation}\label{trwsum}
\Tr_{\frak g} (t^2) = {1 \over 2 h^\vee_{\frak g}} \Tr_{\text{adj}} (t^2) ={1 \over 2 h^\vee_{\frak g}} \sum_{{\alpha \in \Delta_{\frak g} \backslash \Delta_{\frak h}}} \left(\alpha(t)\right)^2~.
\end{equation}
Here the roots in~$\Delta_{\frak h}$ do not contribute, because~$t$ commutes with all elements of~$\frak h$. 

In order to express~$X$ in~\eqref{akx} in terms of more familiar Lie-algebraic data, it is convenient to introduce the vector~$\omega_t \in \frak t^*_{\frak g}$ that is dual to the Cartan element~$t \in \frak t_{\frak g}$,
\begin{equation}\label{omegadefii}
\alpha(t) = \langle \omega_t, \alpha\rangle_{\frak g}~, \qquad \forall \alpha \in \Delta_{\frak g}~,
\end{equation}
and satisfies
\begin{equation}\label{hsq}
\Tr_{\frak g} (t^2) = \langle t,t\rangle_{\frak g} = \langle \omega_t, \omega_t\rangle_{\frak g}~.
\end{equation}
Note that~$\omega_t$ is orthogonal to the hyperplane~$\frak t^*_{\frak h} \subset \frak t^*_{\frak g}$, which is spanned by the roots of~$\frak h$. Recall that the metric~$\langle \cdot, \cdot \rangle_{\frak g}$ is normalized so that the long roots of~$\frak g$ have length-squared~$2$. 

Next, we partition the root system of~$\frak g$ into positive and negative roots~$\Delta^\pm_{\frak g}$. We use~$\Delta_{\frak h}^\pm$ to denote the roots of~$\frak h$ that lie in~$\Delta^\pm_{\frak g}$. The roots~$\alpha \in \Delta^+_{\frak g} \backslash \Delta^+_{\frak h}$ correspond to W-Bosons of positive~$\frak u(1)$ charge~$\alpha(t) > 0$ and transform in a representation of~$\frak h$; the roots~$-\alpha \in \Delta^-_{\frak g} \backslash \Delta^-_{\frak h}$ correspond to W-Bosons of charge~$-\alpha(t)$. Since the Weyl vectors of~$\frak g$ and~$\frak h$ are given by
\begin{equation}\label{rhodef}
\rho_{\frak g} = \half \sum_{\alpha \in \Delta^+_{\frak g}} \alpha~, \qquad \rho_{\frak h} = \half \sum_{\alpha \in \Delta^+_{\frak h}} \alpha~,
\end{equation}
we can rewrite
\begin{equation}\label{srew}
\sum_ {\alpha \in \Delta_{\frak g} \backslash \Delta_{\frak h}} |\alpha(t)| = 2 \sum_ {\alpha \in \Delta^+_{\frak g} \backslash \Delta^+_{\frak h}} \alpha(t) = 4 \langle \omega_t, \rho_{\frak g} - \rho_{\frak h}\rangle_{\frak g}~.
\end{equation}
Substituting~\eqref{hsq} and~\eqref{srew} into~\eqref{akx}, we find that
\begin{equation}\label{xsimp}
X = {16 \langle \omega_t, \rho_{\frak g} - \rho_{\frak h}\rangle_{\frak g}^2 \over \langle \omega_t, \omega_t\rangle_{\frak g}}~.
\end{equation}

Finally, we use the fact that~$\rho_{\frak g} - \rho_{\frak h}$ is orthogonal to the hyperplane~$\frak t^*_{\frak h} \subset \frak t^*_{\frak g}$ spanned by the roots of~$\frak h$. To see this, recall that the roots in~$\Delta_{\frak g}^+ \backslash  \Delta_{\frak h}^+$ are the positively charged W-Bosons, which transform in a representation of~$\frak h$. It follows that~$\Delta_{\frak g}^+ \backslash  \Delta_{\frak h}^+$ is invariant under the Weyl group of~$\frak h$. Hence, the same is true for~$\rho_{\frak g} - \rho_{\frak h}$, which must therefore be orthogonal to~$\frak t^*_{\frak h}$. Since~$\omega_t$ is orthogonal to the same hyperplane, we conclude that~$\rho_{\frak g} - \rho_{\frak h}$ and~$\omega_t$ are parallel. These properties allow us to reduce~\eqref{xsimp} to
\begin{equation}
X = {16} \, \langle \rho_{\frak g} - \rho_{\frak h}, \rho_{\frak g} - \rho_{\frak h}\rangle_{\frak g} = {16}\Big(\langle \rho_{\frak g},\rho_{\frak g}\rangle_{\frak g} - \langle \rho_{\frak h}, \rho_{\frak h}\rangle_{\frak g}\Big)~.
\end{equation}

We would now like to apply the Freudenthal-de\,Vries strange formula
\begin{equation}\label{strangefmla}
\langle \rho_{\frak g}, \rho_{\frak g}\rangle_{\frak g} = {1 \over 12} h^\vee_{\frak g} d_{\frak g}~,
\end{equation}
and similarly for~$\frak h$. This formula only applies if we use the particular metric for which long roots have length-squared~$2$. However, the long roots of~$\frak h$ in general do not have this property with respect to the metric~$\langle \cdot, \cdot\rangle_{\frak g}$ adapted to~$\frak g$. For a general rank one braking~$\frak g \rightarrow \frak h \oplus \frak u(1)$, we have~$\frak h = \oplus_i \frak h_i$, where the~$\frak h_i$ are compact, simple Lie algebras. Then the metrics~$\langle \cdot, \cdot\rangle_{\frak h_i}$ adapted to~$\frak h_i$ are related to the metric adapted to~$\frak g$ as follows,
\begin{equation}\label{nfactordef}
\langle \cdot,\cdot \rangle_{\frak g} = N^{(i)}_{\frak h \subset \frak g} \, \langle \cdot, \cdot\rangle_{\frak h_i}~.
\end{equation}
The normalization factors~$N^{(i)}_{\frak h \subset g}$ depend on the particular embedding of~$\frak h$ in~$\frak g$. They are determined by the lengths of long roots~$\ell_i$ of~$\frak h_i$ with respect to the metric on~$\frak g$,
\begin{equation}\label{nembeddef}
N^{(i)}_{\frak h \subset \frak g} = \half \langle \ell_i, \ell_i\rangle_{\frak g}~, \qquad \ell_i~\text{a long root of}~\frak h_i \subset \frak g~.
\end{equation} 
If~$\frak g$ is simply laced, then all roots have length-squared~$2$, so that~$N^{(i)}_{\frak h \subset \frak g} = 1$, but in general this is not the case. 

The Weyl vector~$\rho_{\frak h}$ of~$\frak h$ defined in~\eqref{rhodef} is the sum of the mutually orthogonal Weyl vectors~$\rho_{\frak h_i}$ of the~$\frak h_i$. We can therefore use~\eqref{nfactordef} and the strange formula~\eqref{strangefmla} to evaluate
\begin{equation}
\langle \rho_{\frak h}, \rho_{\frak h}\rangle_{\frak g} = \sum_i N^{(i)} _{\frak h \subset g} \; \langle \rho_{\frak h_i}, \rho_{\frak h_i}\rangle_{\frak h_i} = {1 \over 12} \sum_i N^{(i)}_{\frak h \subset \frak g} h^\vee_{\frak h_i} d_{\frak h_i}~.
\end{equation}
Therefore, 
\begin{equation}
X = {4 \over 3 } \bigg(h^\vee_{\frak g}d_{\frak g}-\sum_i N^{(i)}_{\frak h \subset \frak g} h^\vee_{\frak h_i} d_{\frak h_i}\bigg)~.
\end{equation}
Substituting into~\eqref{deltaakii}, we conclude that
\begin{equation}\label{deltakwithn}
\Delta a = {16 \over 7 } \bigg(h^\vee_{\frak g}d_{\frak g}-\sum_i N^{(i)}_{\frak h \subset \frak g} h^\vee_{\frak h_i} d_{\frak h_i}\bigg)~, \qquad \Delta k = h^\vee_{\frak g}d_{\frak g}-\sum_i N^{(i)}_{\frak h \subset \frak g} h^\vee_{\frak h_i} d_{\frak h_i}~.
\end{equation}

\smallskip

\begin{table}[h]
\centering
\begin{tabular}{!{\VRule[1pt]}c!{\VRule[1pt]}c!{\VRule[1pt]}c!{\VRule[1pt]}c!{\VRule[1pt]}}
\specialrule{1.2pt}{0pt}{0pt}
$\mathfrak{g}$ & $a_{\frak g}$ &  $c_{\frak g}$ & $k_{\frak g}$  \rule{0pt}{2.6ex}\rule[-1.4ex]{0pt}{0pt} \\
\specialrule{1.2pt}{0pt}{0pt}
\multirow{2}{*}{$\mathfrak{u}(1)$}& \multirow{2}{*}{$1$} &  \multirow{2}{*}{$1$} & \multirow{2}{*}{$0$}\\
 &  & & \\
\hline
\multirow{2}{*}{$\mathfrak{su}(n)$} & \multirow{2}{*}{$\frac{16}{7}n^{3}-\frac{9}{7}n-1$} & \multirow{2}{*}{$4n^{3}-3n-1$} & \multirow{2}{*}{$n^3-n$}\\
& & & \\
\hline
\multirow{2}{*}{$\mathfrak{so}(2n)$}& \multirow{2}{*}{$\frac{64}{7}n^{3}-\frac{96}{7}n^{2}+\frac{39}{7}n$} &  \multirow{2}{*}{$16n^{3}-24n^{2}+9n$}& \multirow{2}{*}{$4n^3-6n^2+2n$}\\
 & & & \\
\hline
\multirow{2}{*}{$\mathfrak{e}_{6}$}& \multirow{2}{*}{$\frac{15018}{7}$} & \multirow{2}{*}{$3750$}& \multirow{2}{*}{$936$} \\
 & & & \\
\hline
\multirow{2}{*}{$\mathfrak{e}_{7}$}& \multirow{2}{*}{$5479$}  & \multirow{2}{*}{$9583$}& \multirow{2}{*}{$2394$}\\
 &  & &\\
\hline
\multirow{2}{*}{$\mathfrak{e}_{8}$}& \multirow{2}{*}{$\frac{119096}{7}$} & \multirow{2}{*}{$29768$} & \multirow{2}{*}{$7440$}\\
 & & &\\
\specialrule{1.2pt}{0pt}{0pt}
 \end{tabular}

\caption{Weyl and~$\frak{so}(5)_R$ anomalies for all~$(2,0)$ SCFTs~$\CT_{\frak g}$.}
\label{gensp}
\end{table}

In order to obtain formulas for~$a_{\frak g}$ and~$k_{\frak g}$, we can break~$\frak g \rightarrow \frak u(1)^{r_{\frak g}}$ by a sequence of rank one breakings, i.e.~by successively removing nodes from the Dynkin diagram of~$\frak g$,  and apply~\eqref{deltakwithn} at every step. If~$\frak g$ is an ADE Lie algebra then all~$N^{(i)}_{\frak h \subset \frak g} = 1$, so that
\begin{equation}\label{finalak}
a_{\frak g} = {16 \over 7} h^\vee_{\frak g} d_{\frak g} + r_{\frak g}~, \qquad k_{\frak g} = h^\vee_{\frak g} d_{\frak g}~, \qquad \frak g \in \left\{A_n,D_n,E_n\right\}~.
\end{equation}
The second term in~$a_{\frak g}$ arises from the~$r_{\frak g}$ tensor multiplets that remain when ~$\frak g$ is completely broken to the Cartan subalgebra. The answer for~$k_{\frak g}$ is in agreement with the known answer discussed around~\eqref{anomform}, and the one for~$a_{\frak g}$ agrees with expectations from holography~\cite{Henningson:1998gx,Tseytlin:2000sf,Beccaria:2014qea}. In Table~\ref{gensp} we display the values of~$a_{\frak g}$ and~$k_{\frak g}$ computed in~\eqref{finalak}, as well as the conjectured value of the~$c$-anomaly~$c_{\frak g}$ in~\eqref{cresult}, for all ADE Lie algebras~$\frak g$. 

When~$\frak g$ is not simply laced, it is generally not possible to find functions~$a_{\frak g}, k_{\frak g}$ that only depend on~$\frak g$ and satisfy~\eqref{deltakwithn} for all rank one adjoint breaking patterns. We will discuss examples in section~4.3, where we revisit the ADE restriction on~$\frak g$.

\subsection{The~$a$-Theorem for RG Flows with~$(2,0)$ Supersymmetry}

In six dimensions, the conjectured~$a$-theorem~\cite{Cardy:1988cwa} (see also~\cite{Elvang:2012st}) states that the~$a$-anomaly strictly decreases under any unitary RG flow that interpolates between a~$\text{CFT}_{\text{UV}}$ at short distances and a~$\text{CFT}_{\text{IR}}$ at long distances,
\begin{equation}
a_{\text{UV}} > a_{\text{IR}}~.
\end{equation}
Broadly speaking, RG flows fall into two categories: those initiated by deforming the~$\text{CFT}_{\text{UV}}$ using a relevant operator, which breaks conformal invariance explicitly, and those initiated by activating a vev and (partially) moving onto a moduli space of vacua, where conformal invariance is spontaneously broken. 

As was stated in section~1.1, the~$(2,0)$ SCFTs in six dimensions do not possess relevant (or marginal) operators that can be used to deform the Lagrangian while preserving~$(2,0)$ supersymmetry~\cite{ckt}. Therefore, the only possible RG flows that preserve that amount of supersymmetry are the moduli-space flows we have analyzed above. They are induced by adjoint Higgsing~$\frak g \rightarrow \frak h \oplus \frak u(1)^n$ with~$\frak h$ semisimple and~$n \leq r_{\frak g}$. If~$\frak g$ is an ADE Lie algebra, we can use~\eqref{finalak} to evaluate~$a_{\text{UV}} = a_{\frak g}$ and~$a_{\text{IR}} = a_{\frak h} + n$, and hence verify that their difference is positive. Since~$r_{\frak g} = r_{\frak h} + n$, this amounts to the statement that
\begin{equation}\label{deltahd}
h^\vee_{\frak g} d_{\frak g} -
 h^\vee_{\frak h} d_{\frak h} > 0~.
\end{equation}
Using the formula for~$k_{\frak g}$ in~\eqref{finalak}, we see that the same combination governs the change~$\Delta k$ in the~$R$-symmetry anomaly. It is is therefore also monotonic under RG flow~\cite{Intriligator:2000eq}. Similarly, using the conjectured formula~\eqref{cresult} for the~$c$-anomaly shows that~$\Delta c$ is also proportional to the left-hand side of~\eqref{deltahd}, and hence positive. Therefore, the class of RG flows that preserve~$(2,0)$ supersymmetry is not sufficient to single out the~$a$-anomaly as the only monotonically decreasing quantity.\footnote{~An analogous situation occurs for RG flows between four-dimensional~$\CN=4$ SCFTs, since their Weyl anomalies~$a$ and~$c$ are always equal.}

The statement that~$\Delta a > 0$ for any flow can also be understood without using the explicit formula for~$a_{\frak g}$ in~\eqref{finalak}, or assuming that~$\frak g$ is simply laced. For rank one breaking patterns~$\frak g \rightarrow \frak h \oplus \frak u(1)$, it follows from~\eqref{deltaakii} that~$\Delta a \sim b^2$, with a positive, model-independent proportionality factor. As was shown in~\cite{Maxfield:2012aw} and reviewed in section~2.2, this relationship is dictated by supersymmetry.  Since~$b$ can only vanish in free theories (see the discussion around~\eqref{fourd} and in section~3.4), it follows that~$\Delta a > 0$, which establishes the~$a$-theorem for rank one flows. Any adjoint breaking pattern can be obtained as a sequence of rank one breakings, by sequentially removing nodes from the Dynkin diagram of~$\frak g$. Therefore, the conclusion~$\Delta a > 0$ applies to all flows induced by adjoint Higgsing. Similarly, it also follows from~\eqref{deltaakii} that~$\Delta k \sim b^2$, and hence the same argument shows that~$\Delta k > 0$ for all such flows.

\subsection{The ADE Classification Revisited}

In section~3.2 we used properties of the dynamical W-strings that exist on the tensor branch of~$(2,0)$ theories~$\CT_{\frak g}$, and their relation to BPS states in five dimensions, to argue that~$\frak g$ must be a simply-laced Lie algebra. Here we will use the results of section~4.1 to offer an alternative perspective on the ADE restriction that only relies on properties of the massless fields on the six-dimensional tensor branch. Specifically, we will use the fact that the difference between the~$R$-symmetry anomalies of the UV and IR theories satisfies the quantization condition~\eqref{wzwquant}, because it multiplies a WZ term in six dimensions~\cite{Intriligator:2000eq},
\begin{equation}\label{dkquantii}
\Delta k = k_{\frak g} - k_{\frak h} \in 6 \Z~.
\end{equation}
In~\cite{Intriligator:2000eq}, this was interpreted as Dirac quantization for the dynamical W-strings on the tensor branch. The same quantization condition can be obtained by considering a non-dynamical instanton-string for an~$\frak{so}(5)_R$ background gauge field. It was argued in~\cite{Ohmori:2014kda,Intriligator:2014eaa} that such a string acts as a source for the dynamical, self-dual three-form fields on the tensor branch, with charges that are related to the~$R$-symmetry anomaly via a Green-Schwarz mechanism. Requiring these charges to satisfy Dirac quantization -- with appropriate adjustments for self-dual fields, as in~\eqref{6diracpair} -- also leads to~\eqref{dkquantii}. 

In section~4.1 we computed~$\Delta k$ for the rank one  breaking patterns~$\frak g \rightarrow \frak h \oplus \frak u(1)$, where~$\frak g$ is an arbitrary compact, simple Lie algebra. When~$\frak g$ is of ADE type, this formula can be integrated to~\eqref{finalak}. The quantization condition~\eqref{dkquantii} then amounts to the statement that~$h^\vee_{\frak g} d_{\frak g} \in 6 \Z$ for any ADE Lie algebra, which is indeed the case, as emphasized in~\cite{Intriligator:2000eq}. When~$\frak g$ is not simply laced, the formula for~$\Delta k$ (and also~$\Delta a$) in~\eqref{deltakwithn} involves the normalization factors~$N^{(i)}_{\frak h \subset \frak g}$, which generally depend on the way that~$\frak h$ is embedded into~$\frak g$. 

As an example, consider~$\frak g = \frak g_{2}$, which has a maximal subalgebra~$\frak{su}(2)_{\circ} \oplus \frak{su}(2)_{\bullet}$. 
Here the subscripts~$\circ$ and~$\bullet$ emphasize the fact that the two~$\frak{su}(2)$ summands are associated with the long and short roots in the~$\frak g_2$ Dynkin diagram~$\circ \hskip-2pt \equiv\hskip-2pt\bullet\,$. It is possible to break~$\frak g_2 \rightarrow \frak{su}(2) \oplus \frak u(1)$ in two inequivalent ways by adjoint Higgsing: we can either delete the long root~$\circ$ from the Dynkin diagram by choosing~$t = t_\circ$ to be the Cartan generator of~$\frak{su}(2)_\circ$, so that~$\frak h = \frak{su}(2)_\bullet$ is unbroken; or we can delete the short root~$\bullet$ by setting~$t = t_\bullet$ and preserve~$\frak h = \frak{su}(2)_\circ$\,. According to~\eqref{nembeddef}, the normalization factors for the two embeddings are determined by the long roots of the subgroup~$\frak h \subset \frak g_2$,\begin{equation}
N_{\frak{su}(2)_\bullet \subset \frak g_2} = {1 \over 3}~, \qquad N_{\frak{su}(2)_\circ \subset \frak g_2} = 1~.
\end{equation}
Substituting into~\eqref{deltakwithn}, we find 
\begin{equation}\label{g2answers}
\Delta k_{\frak{su}(2)_\bullet \subset \frak g_2} = 4 \cdot 14 - {1 \over 3} \cdot 2 \cdot 3 = 54~, \qquad \Delta k_{\frak{su}(2)_\circ \subset \frak g_2}= 4 \cdot 14 - 2 \cdot 3 = 50~.
\end{equation}
This result can also be obtained by directly evaluating the sums over W-Bosons in~\eqref{akx} and~\eqref{trwsum}, using the fact that the adjoint~$\bf 14$ of~$\frak g_2$ decomposes as follows under~$\frak{su}(2)_\circ \oplus \frak{su}(2)_\bullet$,
\begin{equation}\label{g2dec}
{\bf 14} \rightarrow \left({\bf 1}, {\bf 3}\right) \oplus \left({\bf 3}, {\bf 1}\right) \oplus \left({\bf 2}, {\bf 4}\right)~.
\end{equation}  
Note that~$\Delta k_{\frak{su}(2)_\circ \subset \frak g_2}$ in~\eqref{g2answers} is not divisible by six, i.e.~it does not satisfy the quantization condition~\eqref{dkquantii}. This rules out~$(2,0)$ theories~$\CT_{\frak g}$ with~$\frak g = \frak g_2$.

Similar phenomena occur for all non-simply-laced Lie algebras, since they contain roots of different lengths. In general, the subgroup~$\frak h$ decomposes into several compact, simple summands, which give rise to different normalization factors~\eqref{nembeddef} that must be added according to~\eqref{deltakwithn}.\footnote{~For example, we can break~$\frak{so}(7)\rightarrow \frak{su}(2)_\circ \oplus \frak{su}(2)_\bullet \oplus \frak u(1)$ by deleting the middle node from the~$\frak{so}(7)$ Dynkin diagram~$\circ\hskip-2pt-\hskip-2pt\circ\hskip-2.5pt=\hskip-2.5pt\bullet$\,. According to~\eqref{nembeddef}, the normalization factors for~$\frak{su}(2)_{\circ}$ and~$\frak{su}(2)_\bullet$ are~$N_{\circ} = 1$ and~$N_{\bullet} = \half$, respectively. Therefore~\eqref{deltakwithn} gives~$\Delta k = 5 \cdot 21 - 1 \cdot 2 \cdot 3 - \half \cdot 2 \cdot 3 = 96$.} The quantization condition~\eqref{dkquantii} also rules out all other non-ADE Lie algebras~$\frak g$. In order to show this, it suffices to rule out~$\frak g = \frak{so}(5) = \frak{sp}(4)$, since it can be reached from all non-simply-laced Lie algebras other than~$\frak g_2$ by adjoint Higgsing. We can break~$\frak{so}(5) \rightarrow \frak{su}(2)_\bullet \oplus \frak u(1)$ by deleting the long root~$\circ$ from the~$\frak{so}(5)$ Dynkin diagram~$\circ \hskip-3pt =\hskip-3pt\bullet\,$. Substituting the normalization factor~$N_{\frak{su}(2)_\bullet \subset \frak{so}(5)} = \half$ from~\eqref{nembeddef} into~\eqref{deltakwithn} then gives~$\Delta k_{\frak{su}(2)_\bullet \subset \frak{so}(5)} = 3 \cdot 10 - \half \cdot 2 \cdot 3 = 27$, which does not satisfy~\eqref{dkquantii}. 

In summary, the fact that there are no~$(2,0)$ SCFTs~$\CT_{\frak g}$ unless~$\frak g$ is a simply-laced Lie algebra is required by the consistency of the low-energy effective theory on the six-dimensional tensor branch, due to the quantization condition~\eqref{dkquantii}.
 
\section{Compactification to Four Dimensions}

In this section we will consider~$(2,0)$ SCFTs~$\CT_{\frak g}$ on~$\R^{3,1} \times T^2$. Here~$T^2 = S^1_R \times S^1_r$ is a rectangular torus of area~$A = R r$ and modular parameter~$\tau = i \tau_2 = i\left({r \over R}\right)$. We describe their moduli spaces of vacua, which depend on a choice of gauge group~$G$, and the singular points at which interacting~$\CN=4$ theories reside. In addition to the familiar theory with gauge group~$G$ at the origin there are typically additional singular points at finite distance~$\sim A^{-\half}$ (sometimes with a different gauge group), which recede to infinite distance when the torus shrinks to zero size, $A\rightarrow 0$. We use non-renormalization theorems to determine the four-derivative terms in the Coulomb-branch effective action via a one-loop calculation in five-dimensional~$\CN=2$ Yang-Mills theory, which now includes a sum over KK modes, and interpret the result. Many statements in this section have five-dimensional analogues, which were discussed in section~3. We will therefore be brief, focusing on those aspects that are particular to four dimensions. 

\subsection{Two-Derivative Terms and Singular Points on the Coulomb Branch}

As in five dimensions, the two-derivative theory on the Coulomb branch of the toroidally compactified theory is completely rigid, due to the constraints of maximal supersymmetry. The geometry of the moduli space can therefore be understood by first compactifying to five-dimensional~$\CN=2$ Yang-Mills theory on~$S^1_R$, and then analyzing the classical vacua of this Yang-Mills theory on~$\R^{3,1} \times S^1_r$. (Note that the order in which we compactify on the two circles selects an S-duality frame in four dimensions.) Many other aspects of toroidally compactified~$(2,0)$ theories can also be understood by studying the five-dimensional Yang-Mills theory on~$\R^{3,1} \times S^1_r$ (see for instance~\cite{Tachikawa:2011ch} and references therein). In section~5.2 we will follow this logic to determine the four-derivative terms on the Coulomb branch.

As in previous sections, we will focus on rank one Coulomb branches described by a single Abelian vector multiplet, which is associated with a Cartan generator~$t \in \frak t_{\frak g}$ of the gauge algebra~$\frak g$ in five dimensions. In addition to the fields that are already present in five dimensions, the four-dimensional effective theory contains an additional real scalar~$\sigma$, which is the Wilson line of the five-dimensional Abelian gauge field~$a_\mu dx^\mu$ around~$S^1_r$,
\begin{equation}\label{sigmadef}
\sigma = {1 \over 2 \pi r} \int_{S^1_r} a_\mu dx^\mu~.
\end{equation}
With this normalization~$\sigma$ has dimension one, and its kinetic terms agree with those of the other five scalars~$\varphi^I$. Explicitly, we can reduce the five-dimensional kinetic terms in~\eqref{kt}  to zero modes of along~$S^1_r$ (taking into account factors of~$2 \pi r$) and use the relation~$g^2 = 4 \pi^2 R$ from~\eqref{gaugeagain} to obtain the kinetic terms on the four-dimensional Coulomb branch,
\begin{equation}
- {\tau_2 \over 4 \pi} \Tr_{\frak g} (t^2) \left(f \wedge * f + \d_\mu \sigma \d^\mu \sigma + \sum_{I = 1}^5 \d_\mu \varphi^I \d^\mu \varphi^I\right) + \left(\text{Fermions}\right) \subset \SL^{(4)}_{\text{Coulomb}}~,\;\;\tau_2 = {R\over r}~.
\end{equation}

The scalar~$\sigma$ is periodic, because it is the holonomy of the Abelian gauge field~$a_\mu dx^\mu$ in five dimensions, which is in turn embedded into a non-Abelian gauge field, as in~\eqref{nonabeemb}. We will parametrize the periodicity of~$\sigma$ as follows,
\begin{equation}\label{sigmaperdef}
\sigma \sim \sigma + {p \over r}~, \qquad (p >0)~.
\end{equation}
The dimensionless constant~$p$ depends on the choice of gauge group~$G$ in five dimensions, specifically its maximal torus. It is defined to be the smallest positive number that satisfies
\begin{equation}\label{pfind}
\exp\left(2 \pi i \, p\, t\right) = \1_G~,
\end{equation}
where~$\1_G$ denotes the identity element of the Lie group~$G$. Since~$\sigma$ is periodic while the~$\varphi^I$ are not, the~$R$-symmetry is typically~$\frak{so}(5)_R$, as in five dimensions. If~$r\rightarrow 0$, so that the area~$A = R r$ of the torus vanishes, the periodicity of~$\sigma$ in~\eqref{sigmaperdef} disappears. In this limit we obtain a genuinely four-dimensional~$\CN=4$ theory with an accidental~$\frak{so}(6)_R$ symmetry under which the six scalars~$\sigma, \varphi^I$ transform as a vector. 

We will now explore the structure of the Coulomb branch parametrized by the scalars~$\sigma, \varphi^I$. As in higher-dimensions, it is convenient to use the radial variable~$\psi = \left(\sum_I \varphi^I\varphi^I\right)^\half$. For generic~$\langle\sigma\rangle, \langle\psi\rangle$, the gauge symmetry is broken to the commutant~$\frak h \oplus \frak u(1)$ of the Cartan generator~$t$ in~$\frak g$. At the origin~$\langle \sigma\rangle = \langle \psi \rangle = 0$, the gauge symmetry is restored to~$\frak g$. Interestingly, the circle of vacua parametrized by~$\langle \sigma \rangle \neq 0$ and~$\langle \psi\rangle = 0$ typically contains other points at which the gauge symmetry is enhanced beyond~$\frak h \oplus \frak u(1)$. These can be found by examining the commutant of the Wilson line parametrized by~$\langle \sigma\rangle$ inside the gauge group~$G$,
\begin{equation}
\exp\left(2 \pi i \langle \sigma \rangle\, r \, t\right) \in G~.
\end{equation}

A complementary approach to identifying singular points on the Coulomb branch is to track the masses of~$\half$-BPS W-Bosons and their KK modes on~$S^1_r$ as functions of the vevs~$\langle \sigma\rangle, \langle \psi\rangle$. The BPS mass formula for a W-Boson corresponding to a root~$\alpha \in \Delta_{\frak g} \backslash \Delta_{\frak h}$ with KK momentum~$\ell \over r$ takes the following form (see e.g.~\cite{Tachikawa:2011ch}),
\begin{equation}
m^2_W(\alpha, \ell) = \alpha(t)^2 \langle\psi\rangle^2 + \left({\ell \over r} - \langle \sigma\rangle \alpha(t)\right)^2~, \qquad \ell \in \Z~. \label{wkk}
\end{equation}
Just as for the two-derivative terms, maximal supersymmetry ensures that this classical formula is quantum-mechanically exact. At the origin of the Coulomb branch, the~$\ell = 0$ KK modes of all W-Bosons become massless, so that the gauge symmetry is enhanced to~$\frak g$. However, if~$\langle \psi \rangle = 0$, there can be points~$\langle \sigma \rangle \neq 0$ at which certain W-Bosons with~$\ell \neq 0$ become massless, which also enhances the gauge symmetry. Rather than describing these phenomena in full generality, we will illustrate them in several representative examples.

\bigskip

\noindent {\it The Moduli Space for~$\frak g = \frak{su}(n)$}

\medskip

The simply connected Lie group with Lie algebra~$\frak{su}(n)$ is~$SU(n)$, whose center is~$\Z_n$. The possible choices of gauge group are then given by
\begin{equation}
G_j=SU(n)/\mathbb{Z}_{j}~.
\end{equation}
Here~$j$ is a divisor of~$n$, so that the center of~$G_j$ is~$\Z_n/\Z_j$. We will consider the rank one breaking pattern~$\frak{su}(n)\rightarrow  \frak{su}(k) \times \frak{su}(n-k)\times \frak{u}(1)$. In the fundamental representation, a Cartan element~$t$ that leads to this breaking pattern is given by
\begin{equation}
t=\left(\begin{array}{c|c}(n-k) \, \1_{k} & 0 \\ \hline0 & -k \,\1_{n-k} \end{array}\right)~.
\end{equation}
We can now use~\eqref{pfind} to find the periodicity~$p$ of the compact scalar~$\sigma$,
\begin{equation}\label{sunkp}
p= {1 \over j \cdot  \mathrm{gcd}(n-k,k)}~.
\end{equation}
Whenever the Wilson line~$\exp(2 \pi i \langle \sigma \rangle \, r \, t)$ is in the center~$\Z_n/\Z_j$ of the gauge group, the full~$G_j$ gauge symmetry is restored. This occurs precisely when
\begin{equation}\label{siglsolve}
\langle \sigma \rangle = {\ell \over n r}~, \qquad \ell \in \Z~, \qquad 0 \leq \ell < {n \over j \cdot \mathrm{gcd}(n-k,k)}~.
\end{equation}
Here the restriction on the integer~$\ell$ is due to the periodicity of~$\sigma$ dictated by~\eqref{sunkp}. 

We can compare these results with the W-Boson mass formula in~\eqref{wkk}. Under the subalgebra~$ \frak{su}(k) \times \frak{su}(n-k)\times \frak{u}(1)$, the W-Bosons transform as follows,
\begin{equation}
\Big(\,\yng(1)~,\overline{\yng(1)}\,\Big)_{n} \oplus \left(\text{complex conjugate}\right)~,
\end{equation}
where the subscript indicates that the~W-Bosons have~$\frak{u}(1)$ charges~$\pm n$. At the values of~$\langle \sigma \rangle$ in~\eqref{siglsolve} (and for~$\langle \psi \rangle = 0$) the mass formula~\eqref{wkk} predicts that the~$\ell^{\text{th}}$ KK modes of all W-Bosons become massless and can therefore restore the full gauge symmetry~$G_j$. 

The preceding discussion illustrates the fact that there are in general multiple singular points on the~$\langle \sigma\rangle$ circle, which correspond to~$\CN=4$ Yang-Mills theories with enhanced gauge symmetry. This phenomenon even occurs on the Coulomb branch of the simplest rank one theory with gauge algebra~$\frak g = \frak{su}(2)$, where~$\frak{su}(2) \rightarrow \frak u(1)$. In the notation above, this corresponds to~$n = 2$ and~$k=1$.    If the global form of the gauge group is~$SU(2)$, the periodicity of~$\sigma$ in~\eqref{sunkp} is~$p =1$ and according to~\eqref{siglsolve} there are two distinct points, $\langle \sigma\rangle = 0$ and~$\langle \sigma\rangle =  {1 \over 2 r}$, where the~$SU(2)$ gauge symmetry is restored. By contrast, if the gauge group is~$SU(2)/\Z_2 = SO(3)$, then~$p = \half$ and only the point~$\langle \sigma\rangle = 0$ has an enhanced~$SO(3)$ gauge symmetry. 

In the language of~\cite{Kapustin:2014gua, Gaiotto:2014kfa}, the~$SU(2)$ gauge theory in five-dimensions has a~$\Z_2$ one-form global symmetry, which shifts the gauge field by a flat~$\Z_2$ connection. This acts as a global~$\Z_2$ symmetry on the scalar~$\sigma$ defined in~\eqref{sigmadef}, which interchanges the two vacua at~$\langle \sigma\rangle = 0$ and~$\langle \sigma\rangle =  {1 \over 2 r}$. When the gauge group is~$SO(3)$, this~$\Z_2$ global symmetry is gauged and the two vacua are identified.  In the limit~$r\rightarrow 0$, the $\sigma$-circle decompactifies and the points with enhanced gauge symmetry are separated by an infinite distance in moduli space. The importance of such global issues for gauge theories on a circle was recently emphasized in~\cite{Aharony:2013dha,Aharony:2013kma}. Subtleties of the zero-area limit were discussed in~\cite{Gaiotto:2011xs}.

\bigskip

\noindent {\it The Moduli Space for~$\frak g = \frak{so}(2n)$}

\medskip

As our second example, we consider~$\frak{g}=\frak{so}(2n)$, which manifests new phenomena. Here we limit the choice of gauge group to~$G = SO(2n)$, i.e.~the standard group of special orthogonal matrices with~$\mathbb{Z}_{2}$ fundamental group. We consider the breaking pattern~$\frak{so}(2n)\rightarrow  \frak{su}(k)\oplus\frak{so}(2(n-k))\oplus \frak{u}(1)$. In the fundamental representation, a Cartan generator~$t$ that gives rise to this breaking is given by
\begin{equation}
t= \left(\begin{array}{c|c}\begin{array}{c|c} 0 & i \1_{k} \\ \hline -i \1_{k} & 0     \end{array} & 0 \\ \hline 0 & 0_{2n-2k}\end{array}\right)~.
\end{equation}
We can then evaluate the Wilson line,
\begin{equation}\label{wilsonlineso}
\exp\left(2 \pi i \langle \sigma \rangle \, r \, t \right) = \left(\begin{array}{c|c}\begin{array}{c|c} \cos\left(2 \pi \langle \sigma \rangle r\right) \1_k & - \sin\left(2 \pi \langle \sigma \rangle r\right) \1_{k} \\ \hline \sin\left(2 \pi \langle \sigma \rangle r\right) \1_{k} & \cos\left(2 \pi \langle \sigma \rangle r\right) \1_k \end{array} & 0 \\ \hline 0 & \1_{2n-2k}\end{array}\right)~.
\end{equation}
This shows that the periodicity of~$\sigma$ is~$p = 1$. At~$\langle \sigma \rangle = 0$, the full~$SO(2n)$ gauge symmetry is restored. However, the vacuum at~$\langle \sigma \rangle = {1 \over 2 r}$ is also special. At this point, the Wilson line in~\eqref{wilsonlineso} reduces to~$\text{diag}(-\1_{2k}, \1_{2n - 2k})$, so that the gauge symmetry is enhanced to~$\frak{so}(2k) \oplus \frak{so}(2(n-k))$. Note that this gauge algebra cannot be reached from~$\frak{so}(2n)$ by standard adjoint Higgsing using the five non-compact scalars~$\varphi^I$. 

These conclusions are again reflected in the spectrum of W-Bosons. They transform as follows under~$\frak{su}(k) \oplus \frak{so}(2(n-k))\oplus \frak{u}(1)$,
\begin{equation}
\Bigg(\, \yng(1,1)~, \mathbf{1} \,\Bigg)_{2} \oplus \Big(\,\yng(1)~, \yng(1)\, \Big)_{1} \oplus \left(\text{complex conjugate}\right)~, \label{wspecso}
\end{equation} 
where the subscripts denote the~$\frak{u}(1)$ charges with respect to~$t$. The mass formula~\eqref{wkk} shows that the~$\ell = 0$ KK modes of all W-Bosons are massless at the origin, where the gauge symmetry is enhanced to~$SO(2n)$. At~$\langle \sigma \rangle = {1\over 2r}$ and~$\langle \psi \rangle = 0$, the~$\ell=1$ KK mode from the~$\bigg(\,\yng(1,1)~, \mathbf{1} \,\bigg)_{2}$ representation and the~$\ell = -1$ KK mode from its complex conjugate representation are massless, while all other W-Boson modes are massive. Together with the massless gauge Bosons of~$\frak{su}(k) \oplus \frak{so}(2(n-k))\oplus \frak{u}(1)$, they precisely fill out the adjoint representation of~$\frak{so}(2k) \oplus \frak{so}(2(n-k))$, which is the unbroken gauge algebra at that point. Therefore we can obtain an~$\CN=4$ theory with this gauge symmetry by taking the~$r \rightarrow 0$ limit while tuning~$\langle \sigma \rangle = {1 \over 2 r}$. In this limit, the conventional vacuum at~$\langle \sigma \rangle = 0$ moves off to infinite distance. 

\subsection{Four-Derivative Terms in Four Dimensions}

We will now use non-renormalization theorems, and our understanding of the five-dimensional theory from section~3, to determine the moduli dependence of the leading higher-derivative interactions on the Coulomb branch of toroidally compactified~$(2,0)$ theories. As in previous sections, we consider rank one Coulomb branches associated with a particular Cartan element~$t$, which breaks~$\frak g \rightarrow \frak h \oplus \frak u(1)$. We focus on the coefficient function~$f_4(\varphi^I, \sigma)$ of the four-derivative term~$(\d \psi)^4$ in the effective action, 
\begin{equation}\label{f4def4d}
A^{2}f_4(\varphi^I, \sigma)(\partial \psi)^{4} \subset \SL^{(4)}_{\text{Coulomb}}~, \qquad A = R r~,
\end{equation}
but as before, a similar discussion applies to all four-derivative terms in~$\SL^{(4)}_{\text{Coulomb}}$. We will fix the function~$f_4(\varphi^I, \sigma)$ by imposing the following constraints:

\begin{itemize}
\item It is invariant under~$\frak{so}(5)_R$ rotations of the~$\varphi^I$, i.e.~it only depends on the~$\varphi^I$ through the radial variable~$\psi$.

\item It is periodic in~$\sigma$ with period~\eqref{sigmaperdef}, i.e.~it is invariant under~$\sigma \sim \sigma + {p \over r}$. 

\item It is dimensionless. Since the six-dimensional theory is scale invariant, this fixes
\begin{equation}
f_4=f_4\left(r \psi, r \sigma, \tau_2\right)~, \qquad \tau_2 = {r \over R}~.
\end{equation}

\item As in sections~2 and~3, it is harmonic, due to the non-renormalization theorems of~\cite{Paban:1998ea,Paban:1998mp,Paban:1998qy,Sethi:1999qv,Maxfield:2012aw}.   

\item In appropriate regimes, it matches onto the six- and five-dimensional results discussed in  sections~2 and~3. In particular, for values of~$\psi$ that are much larger than any other length scale, it must decay to zero. 

\end{itemize}

The periodicity of~$\sigma$ is conveniently taken into account by expanding~$f_4$ in Fourier modes,
\begin{equation}\label{sigmaft}
f_4\left(r \psi, r \sigma, \tau_2 \right)=\sum_{n \in \mathbb{Z}}f^{(n)}_4\left(r \psi, \tau_2 \right)\exp\left(\frac{2\pi i n r \sigma}{p}\right)~.
\end{equation}
Since the non-renormalization theorem states that~$f_4$ is harmonic, each mode function satisfies
\begin{equation}
\frac{d^{2}}{d(r\psi)^{2}}f^{(n)}_4\left(r\psi,\tau_2\right)+\left(\frac{4}{r\psi}\right)\frac{d}{d(r\psi)}f^{(n)}_4\left(r\psi,\tau_2\right)=\frac{4\pi^{2}n^{2}}{p^{2}}f^{(n)}_4\left(r\psi,\tau_2\right)~. \label{modesols}
\end{equation}
For each value of~$n$, this second order differential equations has two linearly independent solutions, only one of which decays to zero at large~$\psi$, 
\begin{equation}\label{modefn}
f^{(n)}_4(r\psi,\tau_2)=b_n(\tau_2)\left(\frac{1}{(r\psi)^{3}}+\frac{2\pi |n| }{p(r\psi)^{2}}\right)\exp\left(-\frac{2\pi |n| r \psi}{p}\right)~.
\end{equation}
Together with~\eqref{sigmaft}, this completely fixes the coefficient function~$f_4$ in terms of an infinite set of coefficients~$b_n(\tau_2)$ that only depend on the four-dimensional gauge coupling~$\tau_2 = {r \over R}$\,. 

In order to fix~$b_n(\tau_2)$, we consider the limit~$R \rightarrow 0$ with~$\tau_2$ fixed, in which the five-dimensional~$\CN=2$ Yang-Mills theory becomes arbitrarily  weakly coupled. Therefore, the function~$f_4$ can be computed exactly by integrating out W-Bosons at one loop. This is similar to the five-dimensional computation discussed around~\eqref{b5loop}, except that one of the momentum integrals is replaced by a sum over KK momenta~$\ell \over r$ along~$S^1_r$. Thus,
\begin{equation}
f_4\left(r \psi, r \sigma, \tau_2\right)=\frac{\tau^2_2}{32 \pi^2} \sum_{\alpha \in \Delta_{\frak g} \backslash \Delta_{\frak h}}\sum_{\ell \in \mathbb{Z}}\left((r \psi )^{2}+\left(\frac{\ell}{|\alpha(t)|}+r\sigma \right)^{2}\,\right)^{-2}~. \label{4doneloop}
\end{equation}
The overall normalization can been fixed by matching onto the five-dimensional result~\eqref{b5loop} in the limit~$r\rightarrow \infty$.

It is instructive to rewrite the expression for~$f_4$ in~\eqref{4doneloop} as a Fourier series~\eqref{sigmaft}. This can be done using the Poisson summation formula,
\begin{equation}
\sum_{\ell \in \mathbb{Z}}\frac{1}{\left(y^{2}+(\ell+x)^{2}\right)^{2}}= \frac{\pi}{2}\sum_{n\in \mathbb{Z}}\left(\frac{1}{|y|^{3}}+\frac{2\pi |n|}{y^{2}}\right)\exp\Big(2\pi i nx-2\pi|n| |y|\Big)~.
\end{equation}
If we introduce the function
\begin{equation}
\delta_{\mathbb{Z}}(x) = \begin{cases} 1 &  \mathrm{if} \ x \in \mathbb{Z}~, \\ 0 &  \mathrm{if} \ x \notin \mathbb{Z}~,  \end{cases}
\end{equation}
then the Fourier coefficients in~\eqref{modefn} can be succinctly written as  
\begin{equation}
b^{(4)}_n(\tau_2)= \frac{\tau_2^2}{64 \pi}\sum_{\alpha\in \Delta_{\frak g} \backslash \Delta_{\frak h}}|\alpha(t)|\delta_{\mathbb{Z}}\left(\frac{n}{p|\alpha(t)|}\right)~.
\end{equation}
This shows that many Fourier modes that are consistent with the periodicity of~$\sigma$ are absent. For instance, we found above that choosing the gauge group to be~$G = SU(2)$ leads to~$p|\alpha(t)| = 2$. Therefore, only Fourier modes with even~$n$ contribute. This example illustrates why it is not in general possible to obtain the coefficient function~$f_4$, and in particular its five-dimensional limit, by computing the lowest Fourier coefficient~$b_0(\tau_2)$ in a genuinely four-dimensional~$\CN=4$ theory and then restoring the periodicity of~$\sigma$ by summing over all possible values of~$n$ with equal weight. This is a fortiori the case if the moduli space of the theory on a torus of finite area contains interacting~$\CN=4$ theories with different gauge groups, as was the case for the~$\frak{so}(2n)$ examples discussed in the previous subsection. 

We can now expand the exact coefficient function~\eqref{4doneloop} in the four-dimensional limit of vanishing area, $A\rightarrow 0$. This enables us to determine the leading area-suppressed irrelevant operators that describe the RG flow into a genuinely four-dimensional~$\CN=4$ theory at low energies.  (The case~$\frak g = \frak{su}(2)$ was recently examined in~\cite{Lin:2015ixa}.) As was discussed in section~5.1, the moduli space may in general contain several points at which interacting~$\CN=4$ theories reside. In order to describe the RG flow into one of these theories, we must appropriately tune the vevs to its vicinity as we take~$A\rightarrow 0$, while the other singular points on the moduli space recede to infinite distance. For simplicity, we will only carry out this procedure for the familiar interacting vacuum at the origin of moduli space. 

We take the zero-area limit by letting~$r \rightarrow 0$ at fixed~$\tau_2$. The leading term in~\eqref{4doneloop} comes from the~$\ell = 0$ KK mode, while the first subleading correction is a sum over KK modes with~$\ell \neq 0$, which can be evaluated at~$\psi = \sigma = 0$. Therefore,
\begin{equation}\label{expansion}
A^2 f_4 \rightarrow {n_W \over 32 \pi^2} {1 \over \psi^2 + \sigma^2} + {\pi^2 A^2 \tau_2^2 \over 1440} \sum_{\alpha \in \Delta_{\frak g} \backslash \Delta_{\frak h}} \left(\alpha(t)\right)^4 + \CO(A^4) \qquad \text{as}~~ A \rightarrow 0~,
\end{equation}
Since it follows from~\eqref{trdef} that the sum over W-Bosons can be written as~$2 h^\vee_{\frak g} \Tr_{\frak g}\left(h^4\right)$, the terms in the Coulomb-branch effective Lagrangian~\eqref{f4def4d} that follow from~\eqref{expansion} are given by
\begin{equation}\label{coulexp}
{n_W \over 32 \pi^2} {(\d\psi)^4 \over \psi^2 + \sigma^2} + {\pi^2 h^\vee_{\frak g} A^2 \tau_2^2 \over 720} \Tr_{\frak g} \left(t^4\right) \left(\d\psi\right)^4 + \cdots \subset \SL^{(4)}_{\text{Coulomb}}~.
\end{equation}
The term proportional to~$n_W$ is the expected one-loop contribution due to integrating out W-Bosons in the four-dimensional~$\CN=4$ Yang-Mills theory at the origin of moduli space, as discussed around~\eqref{b5loop} and in Figure~\ref{loopfig}. The second term is non-singular and can be extrapolated to the origin, where it arises from a~$\half$-BPS irrelevant operator analogous to the five-dimensional~$F^4$ operators discussed in~\eqref{F4}. Unlike in five dimensions, where these operators were shown to be absent, the compactification of~$(2,0)$ SCFTs on finite-area tori generates these interactions with a definite, non-zero coefficient, which can be extracted from~\eqref{coulexp}. When written in terms of~$\Tr_{\frak g}$, they are single-trace terms, which amounts to a definite linear combination of single- and double-trace terms in the fundamental representation. Note that both operators in~\eqref{coulexp} are invariant under the accidental~$\frak{so}(6)_R$ symmetry that emerges in the zero-area limit. However, subleading~$\CO(A^4)$ corrections break this symmetry to~$\frak{so}(5)_R$. All of them are single-trace operators when written in terms of~$\Tr_{\frak g}$, and it is straightforward to extract their coefficients by expanding~\eqref{4doneloop} to higher order.

\section*{Acknoweldgements}\noindent We are grateful to K.~Intriligator, Z.~Komargodski, and~N.~Seiberg for helpful exchanges and comments on the manuscript. CC and TD would like to thank K.~Intriligator for many valuable conversations, and for collaboration on related topics. TD would like to thank members of the Rutgers high-energy theory group for hospitality and discussions. CC is supported by a Junior Fellowship at the Harvard Society of Fellows. TD is supported by the Fundamental Laws Initiative of the Center for the Fundamental Laws of Nature at Harvard University, as well as DOE grant DE-SC0007870 and NSF grants PHY-0847457, PHY-1067976, and PHY-1205550. XY is supported by a Sloan Fellowship and a Simons Investigator Award from the Simons Foundation.

\appendix

\section{Supervertices in Six and Five Dimensions}

We summarize the super spinor helicity formalism for Abelian~$(2,0)$ tensor multiplets in six dimensions, and for Abelian~$\CN=2$ vector multiplets in five dimensions. We review a general classification of supervertices, and describe those that are needed to understand the non-renormalization theorems in sections~2 and~3 from the amplitude point of view. 

\subsection{Super Spinor Helicity Formalism for Abelian~$(2,0)$ Tensor Multiplets in 6D}

The null momentum of a massless particle in six dimensions can be written in bispinor notation as follows~\cite{Cheung:2009dc,Dennen:2009vk,Boels:2012ie},
\ie
p_{AB} = \lambda_{A\alpha} \lambda_{B\beta} \epsilon^{\alpha\beta}~, \qquad p^{AB} \equiv {1\over 2} \epsilon^{ABCD} p_{CD} = \widetilde\lambda^A{}_{\dot\alpha }\widetilde\lambda^B{}_{\dot\beta }\epsilon^{\dot\alpha\dot\beta}~.
\fe
Here $A,B=1,\cdots,4$ are chiral spinor indices of the $SO(5,1)$ Lorentz group. The lower index labels components of a left-handed spinor and the upper index labels components of a right-handed spinor. The indices~$\alpha, \alphadot = 1,2$ are left- and right-haded chiral spinor indices of the~$SO(4)$ little group for massless particles in six dimensions, and~$\lambda_{A\alpha}\,, {\widetilde\lambda^A \,}_{\dot\alpha}$ are spinor helicity variables in six dimensions. We will only need the~$\lambda_{A\alpha}$, because all fields in a~$(2,0)$ tensor multiplet are left-handed.

The 16 one-particle states in a tensor multiplet are represented by monomials in a set of four Grassmann variables~$\eta_{\alpha a}$, where $\alpha$ is a left-handed chiral spinor index of the~$SO(4)$ little group, while~$a=1,2$ is a doublet index of an~$\mathfrak{so}(3)_R \subset \mathfrak{so}(5)_R$ subalgebra of the~$R$-symmetry. For instance, the five scalars are represented by the little-group singlets $1,\eta_{\alpha a}\eta_{\beta b} \epsilon^{\alpha\beta},\eta^4$, while the self-dual tensor is represented by~$\eta_{\alpha a}\eta_{\beta b}\epsilon^{ab}$. In this formalism, only an~$\frak{so}(3)_R$ subalgebra of the~$R$-symmetry is manifest. On one-particle states, the~$\frak{so}(5)_R$ generators are realized as follows,
\ie\label{rsym}
& H_a{}^b = \sum_i \left(\eta_{i\alpha a}{\partial\over \partial\eta_{i\alpha b}} - 2{\delta_a}^b \right)~, \quad R_{ab}^+ = \sum_i \eta_{i\alpha a}\eta_{i\beta b} \epsilon^{\alpha\beta}~, \quad R^{-ab} = \sum_i {\partial\over\partial \eta_{i\alpha a}} {\partial\over\partial \eta_{i\beta b}} \epsilon_{\alpha\beta}~.
\fe
Here the index~$i$ labels different external particles (see below).  

When acting on one-particle states, the 16 supercharges can be split into eight supermomenta and eight superderivatives,
\ie\label{split}
{\bf q}_{Aa} =  \lambda_{A\alpha}\eta^\alpha{}_a~, \qquad
{\overline {\bf q}}_{Aa} = \lambda_{A\alpha}{\partial\over \partial \eta_\alpha{}^a}.
\fe
An $n$-point superamplitude is a generating function for the scattering amplitudes of~$n$ particles in the tensor multiplet -- labeled by the super spinor helicity variables~$(\lambda_{iA\alpha}, \eta_{i\alpha a})$ with~$i=1, \cdots, n$ -- that is invariant under the~$SO(4)$ little group associated with each particle, and is annihilated by the total supercharges
\ie\label{totals}
Q_{A a} = \sum_{i=1}^n {\bf q}_{iA a}~, \qquad\overline Q_{Aa} = \sum_{i=1}^n \overline{\bf q}_{iA a}~.
\fe

\subsection{Super Spinor Helicity Formalism for Abelian~${\cal N}=2$ Vector Multiplets in 5D}

The formalism for~$\CN=2$ vector multiplets in five dimensions can be obtained by reducing the six-dimensional formalism described in the previous subsection. Now~$A,B$ are~$SO(4,1)$ Lorentz spinor indices and~$\alpha,\beta$ are spinor indices of the $SO(3)$ little group for massless particles in five dimensions. In bispinor notation, the five-dimensional null momentum~$p_{AB}$ satisfies~$\Omega^{AB} p_{AB}=0$, where $\Omega^{AB}$ is the anti-symmetric invariant form on the spinor representation of~$SO(4,1)$. The spinor helicity variables~$\lambda_{A\alpha}$ are also constrained and satisfy
\ie
p_{AB} = \lambda_{A\alpha }\lambda_{B\beta }\epsilon^{\alpha\beta}~, \qquad \Omega^{AB} \lambda_{A\alpha}\lambda_{B\beta}\epsilon^{\alpha\beta}=0~.
\fe
Just as in six dimensions, the~16 one-particle states in the vector multiplet are represented by monomials in~$\eta_{\alpha a}$. For instance, the three helicity states of the five-dimensional gauge Bosons are represented by~$\eta_{\alpha a}\eta_{\beta b}\epsilon^{ab}$. The~$\mathfrak{so}(5)_R$ generators take the same form as in~\eqref{rsym}, and the supercharges are still split into supermomenta and superderivatives according to~\eqref{split}.

\subsection{Classification of Supervertices}

The supersymmetry Ward identity for the eight supermomenta~$Q_{Aa}$ is easily solved by expressing the amplitude in the following form,\footnote{~In a non-Abelian gauge theory, the three-point vertex  involves six rather than eight powers of~$Q$, due to kinematic constraints. This vertex is absent in the Abelian theory.}
\ie
\delta^{(d)}(P)\delta^8(Q) {\CA}(\lambda_{iA\alpha}, \eta_{i\alpha a})~,
\fe
where the~$d$-dimensional delta function enforces conservation of the total momentum~$P$, and $\delta^8(Q)\equiv \prod_{A,a} Q_{Aa}$ is a Grassmannian delta function in the supermomenta. The remaining Ward identities demand that the function ${\cal A}$ is annihilated by all~$\overline{Q}_{Aa}$ upon imposing conservation of the total momentum and supermomentum. Below, we will omit the momentum-conserving delta function, keeping in mind that the spinor-helicity variables~$\lambda_{iA\alpha}$ are constrained by momentum conservation.

We define a supervertex to be a local superamplitude without poles in the momenta. We consider two classes of supervertices: 

\begin{itemize} 

\item An~$F$-term supervertex with~$n$-points and~$k$-derivatives is obtained from
\ie\label{fv}
\delta^8(Q) \CF(s_{ij})~,
\fe
by an~$R$-symmetry rotation. Here~$\CF(s_{ij})$ is a polynomial in the Lorentz-invariant Mandelstam variables~$s_{ij}=-(p_i+p_j)^2$, which is homogeneous of degree~${k\over 2}-2$. Using the commutation relations between the~$R$-symmetry generators in~\eqref{rsym} and the supercharges,\footnote{~These are given by~$
[R^+_{ab}, Q_{Ac}] = 0$~,~$[R^-_{ab}, Q_{Ac}] =  \overline Q_{A(a}\epsilon_{b)c}$~,~$[R^+_{ab}, \overline Q_{Ac}] =  Q_{A(a}\epsilon_{b)c}$~, and~$[R^-_{ab}, \overline Q_{Ac}] = 0$.} it can be checked that and~$n$-point, four-derivative supervertex of the form~$\delta^8(Q)$ is annihilated by $R^-_{ab}$ and the traceless part of ${H_a}^b$ in \eqref{rsym}. Furthermore, $\delta^8(Q)$ has eigenvalue $4(4-n)$ with respect to the trace ${H_a}^a$. Therefore, $\delta^8(Q)$ is the lowest-weight state in a set of supervertices that transform in the~$[n-4,0]$ representation of $\mathfrak{so}(5)_R$, i.e.~they are symmetric, traceless~$(n-4)$-tensors. Note that a four-point, four-derivative vertex of the form~$\delta^8(Q)$ is an~$\mathfrak{so}(5)_R$ singlet.

\item A~$D$-term supervertex can be written as
\ie\label{dv}
\delta^8(Q) \overline{Q}^8 \CD(\lambda_{iA\alpha},\eta_{i\alpha a})~,
\fe
where $\overline{Q}^8\equiv \prod_{A,a} \overline Q_{Aa}$ is the product of all superderivatives and~$\CD$ is an arbitrary polynomial expression in the super spinor helicity variables. Such vertices automatically solve the supersymmetry Ward identities, and they contain at least eight derivatives. Note that our definitions of~$F$- and~$D$-term vertices allows for the possibility that an~$F$-term supervertex with sufficiently many derivatives may be expressible as a~$D$-term supervertex. 

\end{itemize}

It is a well-supported conjecture that~$F$- and~$D$-term supervertices exhaust the set of local amplitudes that satisfy the supersymmetry Ward identities. While this has not been proven in full generality, it is known to be true in many cases. For~$R$-symmetry singlet supervertices in maximally supersymmetric Yang-Mills theories, it is equivalent to the superspace classification of~\cite{Bossard:2010pk}. For color-ordered supervertices that are not~$R$-symmetry singlets, the statement agrees with the classification of supersymmetric deformations in~\cite{Chang:2014kma}, extending the on-shell superfield method of~\cite{Movshev:2005ei, Movshev:2009ba}. In theories that are superconformal at the two-derivative level, the classification into~$F$- and~$D$-terms is consistent with results from superconformal representation theory~\cite{ckt}. The non-renormalization theorems discussed in this paper only rely on a classification of four-point supervertices, and the absence of six-point supervertices that are~$R$-symmetry singlets, both of which are known results. 

\subsection{Supervertices on Tensor and Coulomb Branches}

We will now explicitly describe the four- and six-point supervertices that arise at four-derivative order on the tensor branch of six-dimensional~$(2,0)$ theories and the Coulomb branch of~$\CN=2$ Yang-Mills theories in five dimensions. 

Consider a coupling of the form~$f_4(\Phi^I) H^4$ in six dimensions. (An analogous discussion applies to terms of the form~$f_4(\phi^I) F^4$ in five dimensions.) Expanding the coefficient function around a vev~$\langle \Phi^I\rangle$, as in~\eqref{fexp} leads to five- and six-point vertices as in~\eqref{f4expand},
\begin{equation}
\partial_I f_4|_{\langle\Phi\rangle} \delta\Phi^I H^4~, \qquad \partial_I\partial_J f_4|_{\langle\Phi\rangle} \, \delta\Phi^I \delta\Phi^J H^4~.
\end{equation}
The supersymmetric completion of the five-point coupling is given by the following five-point supervertex,
\ie\label{paint}
\partial_I f_4|_{\langle\Phi\rangle} \, \hat e^I \cdot \left( \hat n\, \delta^8\Big(\sum_{i=1}^5Q_i\Big) \right)_{\frak{so}(5)_R}~.
\fe
Here~$\hat e^I, \hat n$ are auxiliary vectors in~$\R^5$, which are rotated by~$\frak{so}(5)_R$. The~$\hat e^I$ are unit basis vectors, and~$\hat n$ is an~$\frak{so}(5)_R$ highest weight state. The subscript~$\frak{so}(5)_R$ in~\eqref{paint} indicates an average over all~$R$-symmetry rotations, which act on~$\delta^8(Q)$ according to~\eqref{rsym} and simultaneously rotate the vector~$\hat n$. Since~$\delta^8(\sum_{i=1}^5 Q_i)$ is a lowest-weight state of an~$\frak{so}(5)_R$ vector, this average is non-zero.

The six-point coupling in~\eqref{paint} can be decomposed into an~$R$-symmetry singlet and a symmetric, traceless tensor of rank two. The latter can be completed into a six-point~$F$-term supervertex of the form
\ie\label{svs}
\partial_I\partial_J f_4|_{\langle\Phi\rangle} \, \hat e^I \, \hat e^J \cdot \left( \hat n \, \hat n \delta^8\Big(\sum_{i=1}^6 Q_i\Big) \right)_{\frak{so}(5)_R}.
\fe
Here the bi-vector~$\hat n\hat n$ is a highest weight state of a rank two, symmetric, traceless~$\frak{so}(5)_R$ tensor. Since~$\delta^8(\sum_{i=1}^6 Q_i)$ is a lowest-weight state in the same~$R$-symmetry representation, the~$\frak{so}(5)_R$ average in~\eqref{svs} is non-vanishing. By contrast, the~$R$-symmetry singlet six-point coupling in~\eqref{paint} does not arise from any supervertex. 

\newpage

\bibliographystyle{utphys}
\bibliography{twozeroanomaly}

\end{document}